\documentclass[]{rmaa}

\usepackage{paralist}
\usepackage[T1]{fontenc}

\usepackage{psfrag,color}

\usepackage[latin1]{inputenc}
\usepackage{epsfig}
\usepackage{amsmath}
\usepackage{graphicx}
\usepackage{subfig}
\usepackage{multirow}
\usepackage{array}
\usepackage{rotating}
\usepackage[normalem]{ulem}

\title{New photometry and spectroscopy of DW Cancri} 

\author{
S.H. Ram\'irez\altaffilmark{1},
O. Segura Montero\altaffilmark{1},
R. Michel\altaffilmark{2} and
J. Echevarr\'ia\altaffilmark{1},
}

\altaffiltext{1}{
Instituto de Astronom\'ia, 
Universidad Nacional Aut\'onoma de M\'exico, 
Apartado Postal 70-264, Ciudad Universitaria, 
M\'exico D.F., C.P. 04510, M\'exico}
\altaffiltext{2}{Instituto de Astronom\'ia, 
Universidad Nacional Aut\'onoma de M\'exico, 
Apartado Postal 877, 
Ensenada, Baja California, C.P. 22830 M\'exico}

\shortauthor{Ram\'irez et al.}
\shorttitle{New observations of DW Cnc}

\abstract{ We present new observations of the cataclysmic variable DW~Cancri, after the system recovered from a low state.  We performed a power spectrum analysis that reveals a clear signal of the 38 min spin period in our photometric data. Our spectroscopic power spectrum search was consistent with studies performed before the low state, showing the orbital and spin modulations. We also conducted a Doppler Tomography study which exhibits a disc structure and an enhanced emission region, possibly related to a hot spot component. Through a wavelet transform analysis we also found evidence of the 70~min spin-orbit beat period. We interpret these results as indication of a partial recovery of the system. 
Nonetheless, DW~Cnc does not yet show all the original photometric modulations reported in 2004, before experiencing the low state. Namely, these signals are a 38 min spin period, an 86 min orbital period, a 70 min beat period between the latter modulations, and an unresolved 110 min period. Even though our photometry shows a  spin cycle modulation and a long 108 min period, it lacks the beat and orbital period signals. 
 Thus, we propose further observations of DW Cancri, to elucidate if the  mechanisms giving rise to the signatures observed prior to the low state, require more time to reactivate.
}

\resumen{Presentamos nuevas observaciones de la variable catacl\'ismica DW Cancri despu\'es de que el sistema se ha recobrado de un estado bajo. Realizamos un an\'alisis de espectro de potencias que revela clara se\~nal del per\'iodo de esp\'in de 38 minutos en nuestros datos fotom\'etricos.  Nuestro espectro de potencias de la espectroscop\'ia fue consistente con estudios realizados antes del estado bajo, mostrando las modulaciones de los per\'iodos orbital y de esp\'in. Tambi\'en realizamos un estudio de Tomograf\'ia Doppler que exhibe una estructura de disco y una regi\'on de emisi\'on intensa, posiblemente causada por una zona de impacto (\textit{hot spot}). Mediante un an\'alisis de ond\'iculas encontramos una presencia localizada del per\'iodo de batimiento de 70 minutos. Interpretamos este conjunto de resultados como una indicaci\'on de una recuperaci\'on parcial del estado bajo. Sin embargo, DW Cancri a\'un no muestra las se\~nales originales reportadas en 2004. A saber, estas modulaciones son: un per\'iodo de esp\'in  de 38 minutos, un per\'iodo orbital de 86 minutos, un per\'iodo de batimiento entre las se\~nales anteriores de 70 minutos, y un per\'iodo irresoluto de 110 minutos. Y aunque nuestra fotometr\'ia muestra el ciclo de esp\'in y un per\'iodo largo de 108 minutos, no muestra indicaci\'on de los per\'iodos de batimiento y orbital. Por ende, proponemos observaciones adicionales de DW Cnc, para dilucidar si los mecanismos que activan las modulaciones reportadas antes del estado bajo, requieren de mayor tiempo para reactivarse.
}
\addkeyword{stars: cataclysmic variables, intermediate polars, individual (DW~Cnc) -- techniques: photometric; spectroscopic}
\begin{document}
\maketitle
\section{Introduction}
\label{sec:intro}
As described by the standard model \citep[e.g.][]{warner:1971}
cataclysmic variables (CVs) are semi-detached binary systems consisting of a late-type secondary star filling its critical Roche surface, that transfers matter into a white dwarf (WD) via an accretion disk \citep[see][for a comprehensive review on CVs]{warner:1995}. CVs can be divided  in systems with outbursts, like dwarf novae, and non-outburst systems like intermediate and polar systems \citep[e.g.][]{hel:2001}. Intermediate Polar Systems (IPs) are a class of CVs whose primary component is a magnetic white dwarf. The presence of a moderate magnetic field (0.1-10 MG) in these systems is strong enough to inhibit the formation of the innermost regions of the accretion disk, but is not sufficiently strong to  synchronise the rotation of the white dwarf with the orbit \citep[e.g.][]{hameury:2017}. The number of Intermediate polars is small (about one percent) compared with the total population  of CVs \citep{warner:1995}.

DW Cancri (hereafter DW Cnc)  was identified as a CV by \citet{stepanian:1982}. \citet{rodriguez:2004} performed its first time-resolved spectroscopic study. Their analysis led them to suggest that the system is a short-period intermediate polar below the period-gap, whose photometric data resembles the behaviour of VY Scl stars. They found the Balmer and He I lines to be modulated with two periods: 86.10$\pm$0.05 min, associated with the orbital period;  and 38.58$\pm$0.02 min, likely corresponding to the WD spin period. \citet{patterson:2004} confirmed the intermediate polar nature of DW Cnc. A radial velocity search performed by these authors detected the 86 and 38 min periods. Additionally, in a photometric power search they found a 70 min signal, consistent with the beat period of the binary (compelling evidence of its IP nature), and also a weak periodic signal at 110 min, which was left as an unsolved problem. \citet{nea19} reported positive {\it XMM-Newton} observations in 2012 in the range 0.3 - 10 Kev; their light curves show evidence of a period around 38 min and also a signature around 75 min, both consistent within the errors to the spin and beat periods, respectively.

In a previous publication \citep{segura:2020}, a radial velocity study of DW Cnc was presented, with observations performed during a low state in 2018-2019. Through a power spectrum analysis they found the 86 min signal associated with the orbital period and two much weaker modulations associated with the 70 min beat period and the 38 min spin period. Particularly, the 38 min signal was significantly weaker than that previously published by \citet{rodriguez:2004} and \citet{patterson:2004}. To explain this substantial change \citet{segura:2020} suggested that the sudden drop into a low state -- caused by an episode of low mass transfer from the companion--  inhibited the lighthouse effect produced by the rebound emission, thus rendering the spin period of the WD difficult to detect. Such variability in the behaviour of DW Cnc exemplifies the importance of pursuing  follow-up observations of the system. \par
Hence, in Section \ref{sec:observations} of this paper we present new photometric and spectroscopic observations of DW Cnc. In Section \ref{sec:radvel} we show a radial velocity study of the $H\alpha$ and He I 5876 \AA ~emission lines. We performed a power spectrum search of our photometric and spectroscopic data, which is shown in Section~\ref{powspec}, followed by a wavelet transform analysis in Section~\ref{sec:wavelet}. We also carried out a Doppler Tomography study in Section \ref{sec:tom}. We close the article with a discussion of our results and our conclusions in Sections~\ref{sec:discus} and \ref{sec:conclusions}, respectively.

\section{Observations and Reduction}
\label{sec:observations}

\subsection{Photometry}

CCD photometry was obtained on 2020 March 8-9 and 15-16, with the 0.84m telescope at the Observatorio Astron\'omico Nacional at San Pedro M\'artir, located in Baja California, Mexico. V and R images were obtained sequentially during the four nights with an e2vm E2V-4240 2048x2048 CCD using 2x2 binning. The exposure times were of 30s and 20s for the V and R filters, respectively. Data reduction was carried out with the {\sc iraf}\footnote{ IRAF is distributed by the National Optical Astronomy Observatories, which are operated by the Association of Universities for Research in Astronomy, Inc. (AURA), under cooperative agreement with the National Science Foundation (NSF).} software system. After bias and flat field corrections, aperture photometry of DW Cnc and some field stars was obtained with the {\sc phot} routine. The same comparison star (RA=7:58:58, DEC=+16:15:07) used by \citet{patterson:2004} was adopted, assuming the reported magnitudes of  B=15.89, V=15.21, and R=14.82. The log of photometric observations is shown in Table~\ref{photlog}.

\subsection{Spectroscopy}

\label{sec:obs}
 Spectra were obtained with the 2.1m telescope of the Observatorio Astron\'omico Nacional at San Pedro M\'artir, using the Boller and Chivens spectrograph and a Spectral Instrument CCD detector in the 5500 - 6500~\AA ~range (resolution $\sim$ 1200), on the nights of 2020 Mar 15 and 16. The exposure time for each spectrum was 300~s.   Standard~{\sc iraf}
procedures were used to reduce the data.   The log of spectroscopic observations is shown in Table~\ref{speclog}. The spectra show strong $H\alpha$ $\lambda$6563 \AA ~ and He I $\lambda$5876 \AA ~emission lines. The typical S/N ratio of the individual spectrum is of $\sim$20 for the emission lines. The spectra were not normalized.

\begin{table}
\centering
	\caption{Log of photometric  observations for DW Cnc.}
    \label{photlog}
    \begin{tabular}{lccc}
        \hline
        Date & Julian Date & ~~~~~~~~~~~~~~~~~~~~~~~No. of exposures    \\
               & (2450000 +) &  V    &  R   \\
        \hline
        08  March  2020 & 8916  & 201  & 210 \\
        09  March  2020 & 8917  & 201  & 219 \\
        15  March 2020 & 8923  & 197  & 191 \\
        16  March 2020 & 8924  & 131  & 133 \\

       \hline
\end{tabular}
\end{table}

\begin{table}
\centering
	\caption{Log of spectroscopic observations for DW Cnc.}
    \label{speclog}
    \begin{tabular}{lcc}
        \hline
        Date & Julian Date & ~~~No. of     spectra    \\
               & (2450000 +)           \\
        \hline
15  March 2020 & 8923  & 28  \\
16  March 2020 & 8924  & 51  \\

       \hline
\end{tabular}
\end{table}       

\section{Radial Velocities}
\label{sec:radvel}



The radial velocity of the emission lines in each spectra were computed using the {\sc rvsao} package in {\sc iraf}, with the {\sc convrv} function, constructed by J. Thorstensen (2008, private communication). This routine follows the algorithm described by \citet{schneider:1980}, convolving the emission line with an antisymmetric function, and assigning the zero value of this convolution as the midpoint of the line profile. As in \citet{segura:2020}, we initially used the {\sc gau2} option, available in the routine, which uses a negative and a positive Gaussian to convolve the emission line, and needs the input of the width and separation of the Gaussians. This method traces the emission of
the wings of the line profile, presumably arising from the inner parts of the
accretion disc. \par
Following the methodology described by \citet{shafter:1986}, we made a diagnostic diagram to find the optimal Gaussian separation, by performing a non-linear least-squares fit \citep{lmfit:2014} of a  simple circular orbit to each trial:
\begin{equation}
V(t) = \gamma + K_1 \sin\left(2\pi\frac{t - t_0}{P_{orb}}\right),
\label{radvel}
\end{equation}
where $\gamma$ is the systemic velocity, $K_1$ the semi-amplitude, $t_0$ the time of inferior conjunction of the donor and $P_{orb}$ is the orbital period. We employed $\chi_{\nu}^2$ as our goodness-of-fit parameter. Note that we have fixed the orbital period, as derived in \citet{segura:2020}, and therefore we only fit the other three parameters. 

\begin{figure*} 
    \includegraphics[width=\columnwidth]{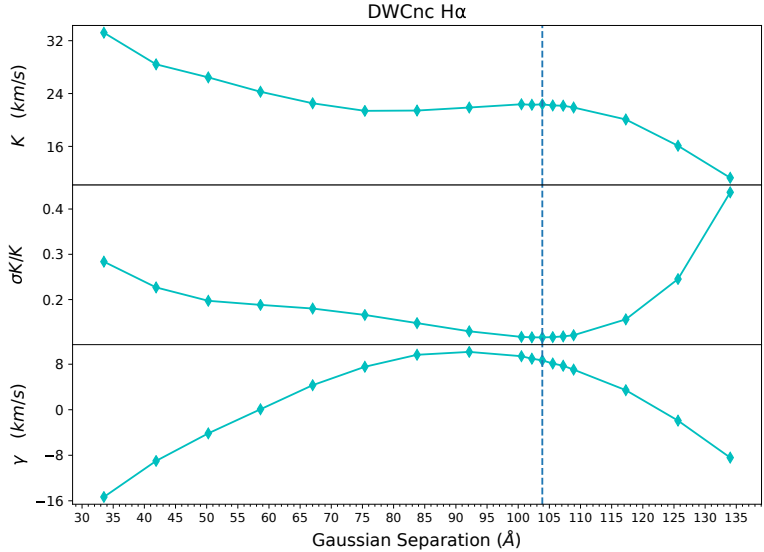}
    \caption{Diagnostic diagram of the $H\alpha$ emission line.  The vertical blue dashed line indicates an optimal separation of 103.9 \AA~(62 pixels). The used width for the Gaussians was 16.8 \AA~(10 pixels). See text for further discussion.}
    \label{fig:diagnostic-ha}
\end{figure*}

\begin{figure*} 
    \includegraphics[width=\columnwidth]{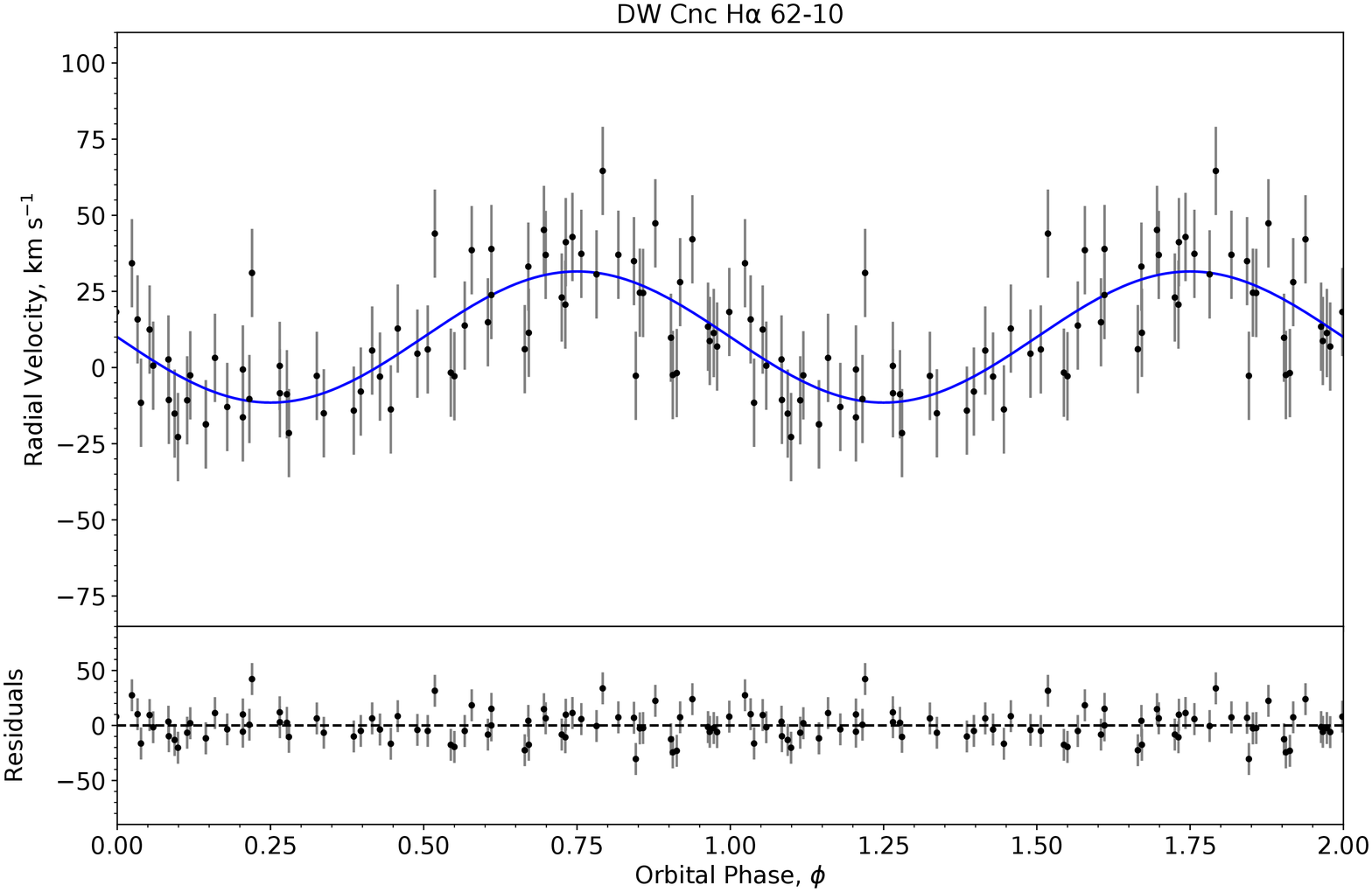}
    \caption{Radial velocity curve for the best solution of the $H\alpha$ emission line. The best fit is shown as the blue line, and the $1\sigma$ error bars have been scaled so that the goodness-of-fit parameter $\chi_{\nu}^2=1$.}
    \label{fig:radvel-ha}
\end{figure*}

\begin{figure*} 
    \includegraphics[width=\columnwidth]{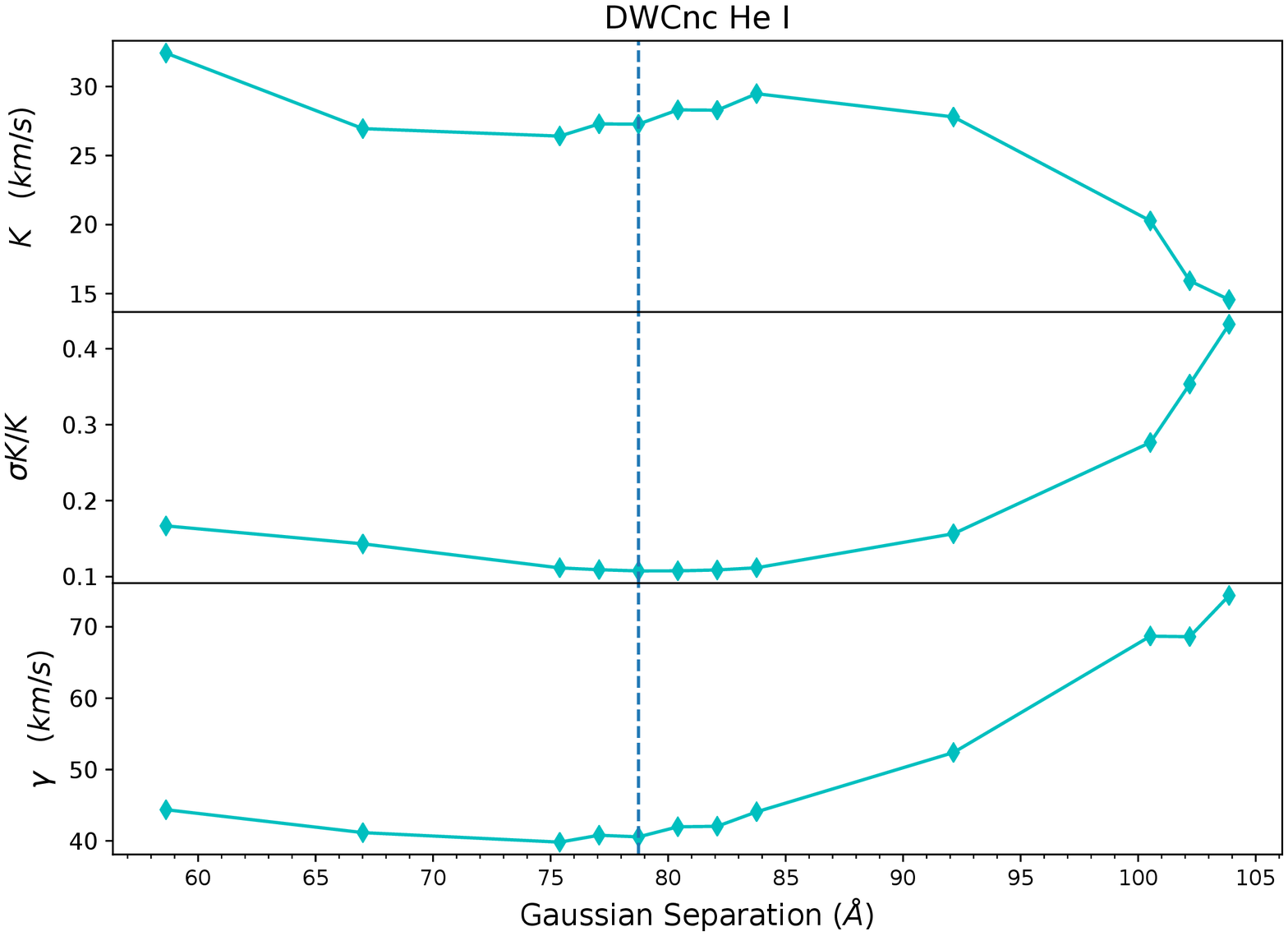}
   
    \caption{Diagnostic diagram of the He I $\lambda5876$ \AA   ~emission line.  The vertical blue dashed line indicates an optimal separation of 78.7 \AA~(47 pixels). The used width for the Gaussians was 11.7 \AA~(7 pixels). See text for further discussion.}
    \label{fig:diagnostic-he}
\end{figure*}

\begin{figure*} 
    \includegraphics[width=\columnwidth]{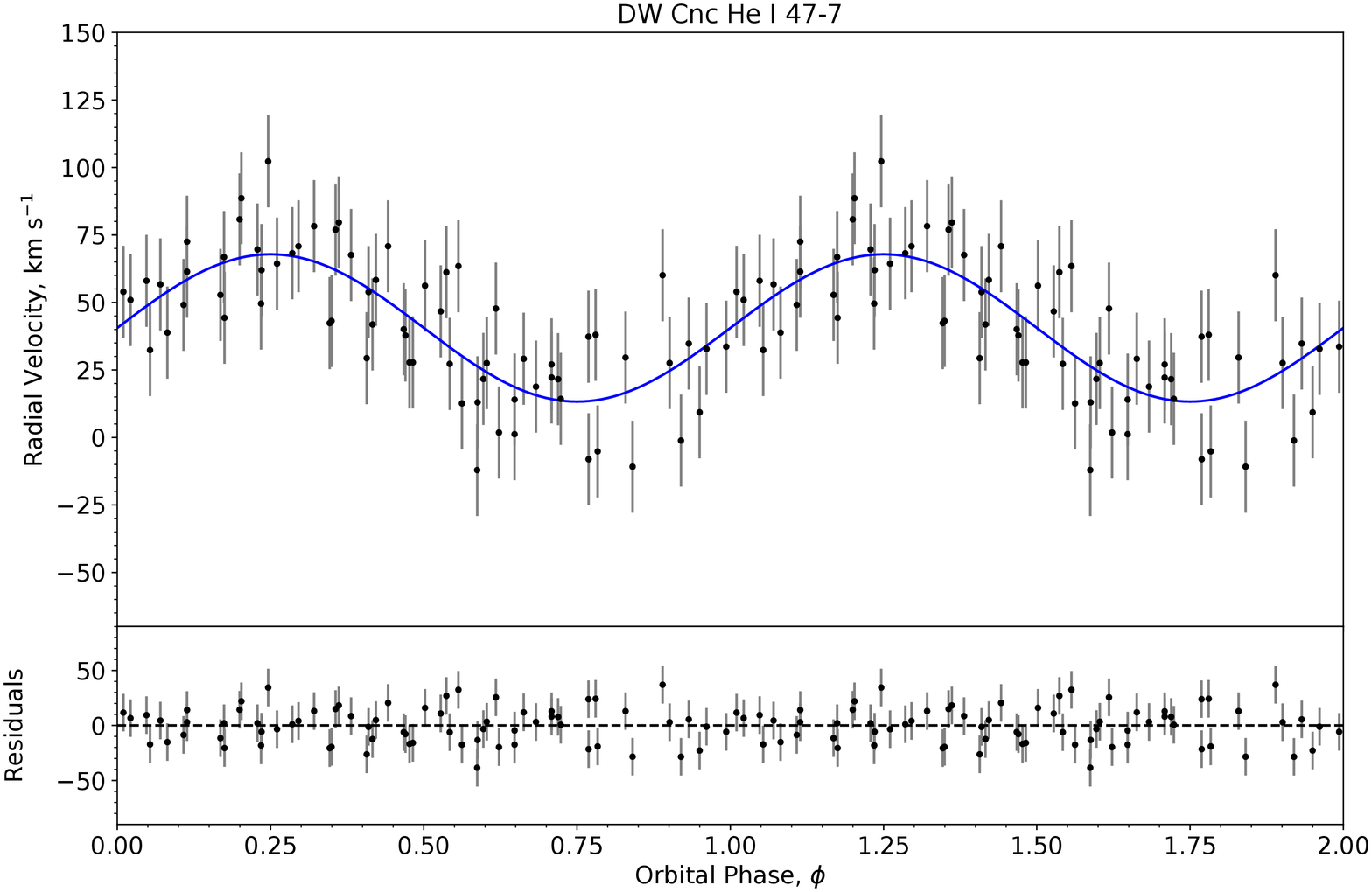}
    \caption{Radial velocity curve for the best solution of the HeI emission line. The best fit is shown as the blue line, and the $1\sigma$ error bars have been scaled so that the goodness-of-fit parameter $\chi_{\nu}^2=1$.}
    \label{fig:radvel-he}
\end{figure*}

In particular, a control parameter is defined in this diagnostic, $\sigma_{K} / K$, whose minimum is a very good indicator of the optimal fit. 
The diagnostic diagram for $H\alpha$ is displayed in Figure~\ref{fig:diagnostic-ha}, while the orbital fit for its best solution is exhibited in Figure~\ref{fig:radvel-ha}. The diagnostic diagram of the He~I~$\lambda5876$~{\AA} emission line is shown in Figure~\ref{fig:diagnostic-he}, and its orbital fit appears in Figure~\ref{fig:radvel-he}. The $1\sigma$ error bars of the radial velocity fits were scaled so that the goodness-of-fit $\chi_{\nu}^2$=1.
The parameters yielded for the optimal orbital fit of both emission lines are shown in Table~\ref{orbpar}, with the respective estimated standard errors for the best-fit values.

Furthermore, as explained in Section~\ref{sec:ps-spec}, following \citet{patterson:2004}, we also implemented the {\sc dgau} convolution option to perform an additional power spectrum. This option uses the derivative of a single Gaussian, and only requires the input of the Gaussian width.\par

\begin{table}

\centering
\caption{Orbital Parameters obtained from H$\alpha$ and He I $\lambda$ 5876, using the wings of the lines.} 
\label{orbpar}
\begin{tabular}{lll}
\hline
Parameter  &  H$\alpha$     &     He I $\lambda 5876$               \\
\hline
  $\gamma$ (km\,s$^{-1}$) & 10 $\pm 2$     & 40
  $\pm 2$   \\
   $K_1$ (km\,s$^{-1}$)   & 21.5 $\pm 2.5$     & 27.3  $\pm 2.9$   \\
  $HJD_0$*          & 0.803 $\pm 0.001$    & 0.833  $\pm 0.001$  \\
  $P_{orb}$  (min)        & Fixed** & Fixed** \\
\hline
\end{tabular}\\
*(24558923+ days)\\
**86.10169~$\pm$~0.00031 min\\
\end{table}

\section{Power spectrum search}
\label{powspec}

We made a power spectrum analysis of the V and R photometric bands,  and a power search of the measured radial velocities of the $H\alpha$ and the He~I~$\lambda$5876~{\AA} emission lines,  using a Lomb-Scargle algorithm \citep{scargle:1982} in both cases. 

For each periodogram, we computed the False-Alarm-Probability (Hereafter FAP), using a function included in \citet{Astropy-Collaboration:2013aa}. As explained by \citet{vanderplas:2018}, the FAP quantifies the significance of a peak, by calculating the probability that the random variations in the data lead to a peak of similar magnitude, conditioned on the assumption of the null hypothesis of having no periodic signal present in the data. Following \citet{deLira:2019}, we identified as significant peaks those whose FAP was less than 0.01, i.e., with a significance level greater than 99 percent. 

\subsection{Photometric data}
\label{sec:photometricdata}
\begin{figure*} 
    \includegraphics[width=\columnwidth, height=6.4cm]{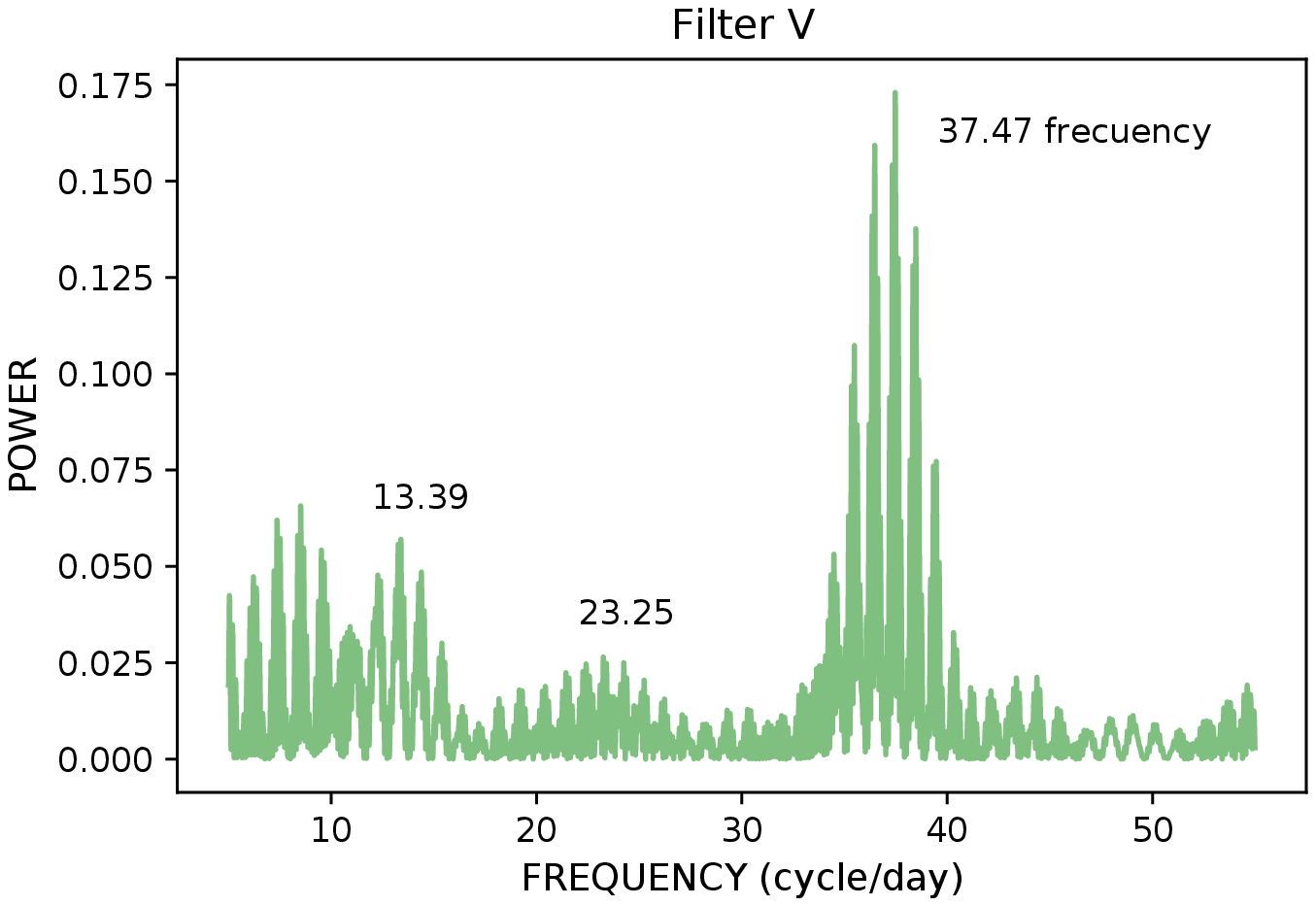}
    \includegraphics[width=\columnwidth, height=6.4cm]{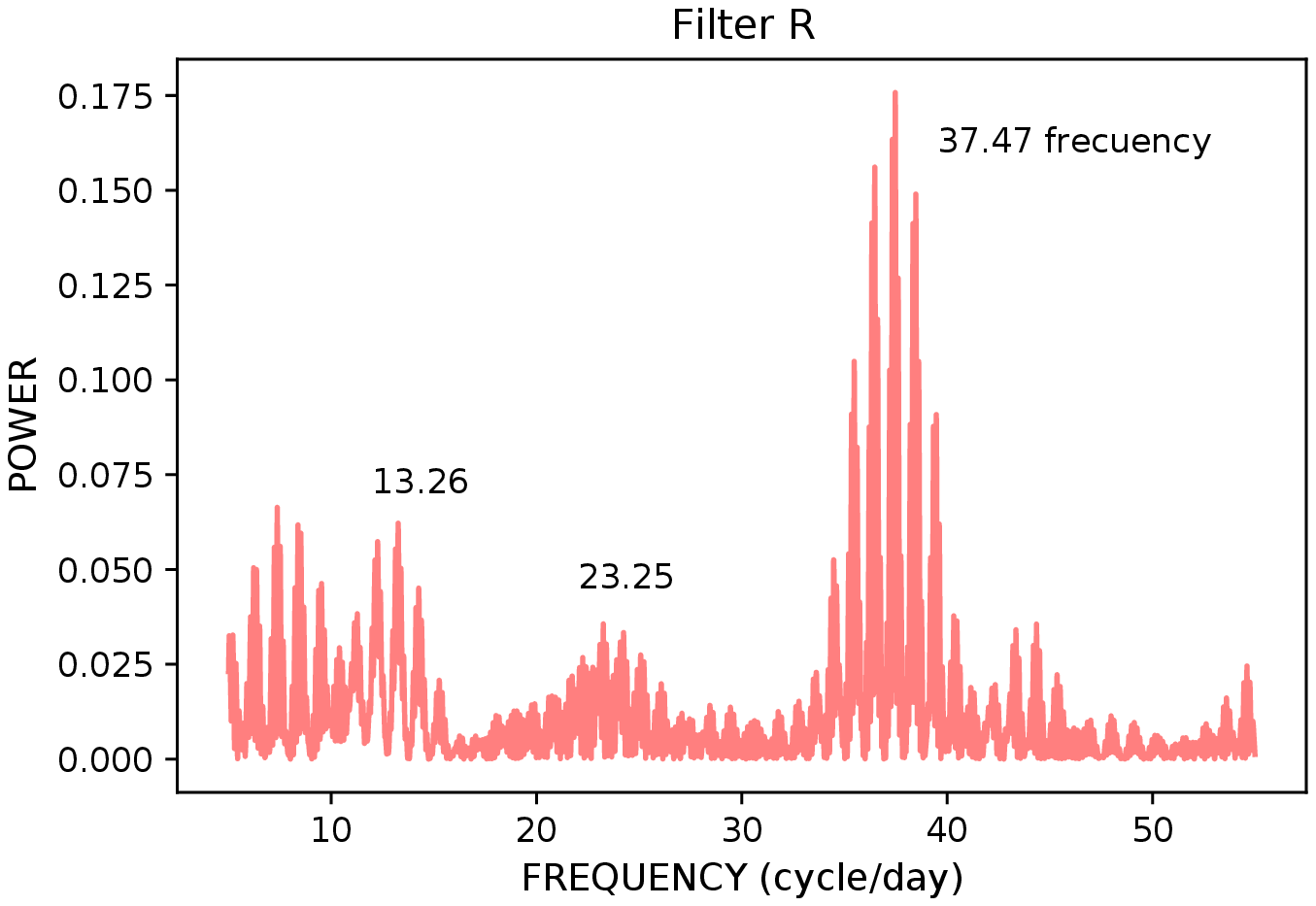}
    \includegraphics[width=\columnwidth, height=6.4cm]{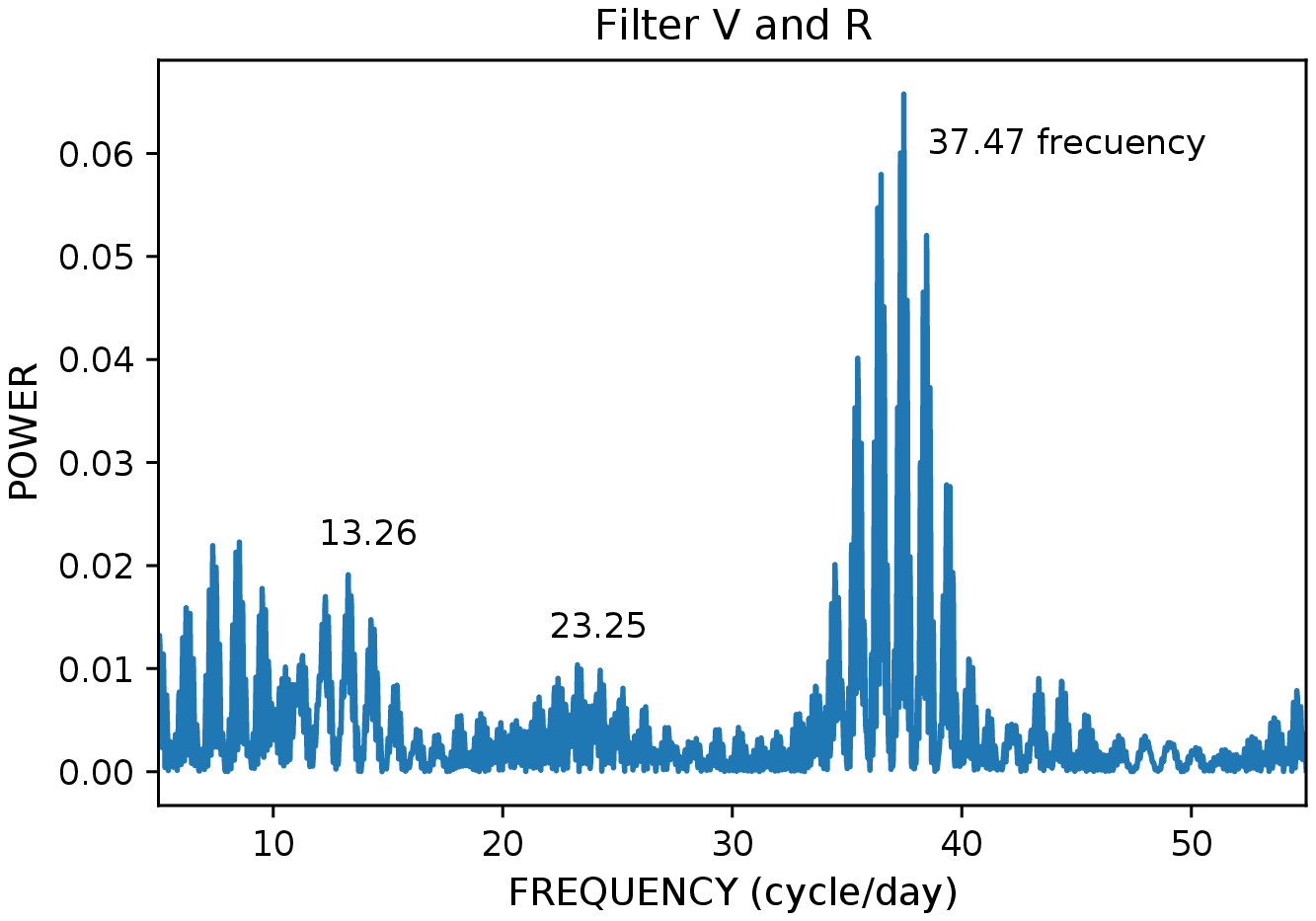}
	\caption{Power spectra of the photometric observations. From top to bottom the panels depict the power spectra of the V band, the R band and of the combined data. See text for further discussion.}
    \label{fig:PS-fotometria}
\end{figure*}

In Figure \ref{fig:PS-fotometria} we show the results of the photometric frequency analysis. The results from the V band  are in the upper panel, where we observe a prominent peak at a frequency of 37.47 cycles/day (associated with the 38 min spin period). Two other weak signals are also present: 13.26 and 23.25~cycles/day, which correspond to periods of 108 min and 62 min, respectively. We do not find the orbital period, and contrary from \citet{patterson:2004} and \citet{nea19}, our periodogram does not detect the 70 min spin-orbit beat period.\par
\begin{figure*} 
    \includegraphics[width=\columnwidth, height=7.5cm]{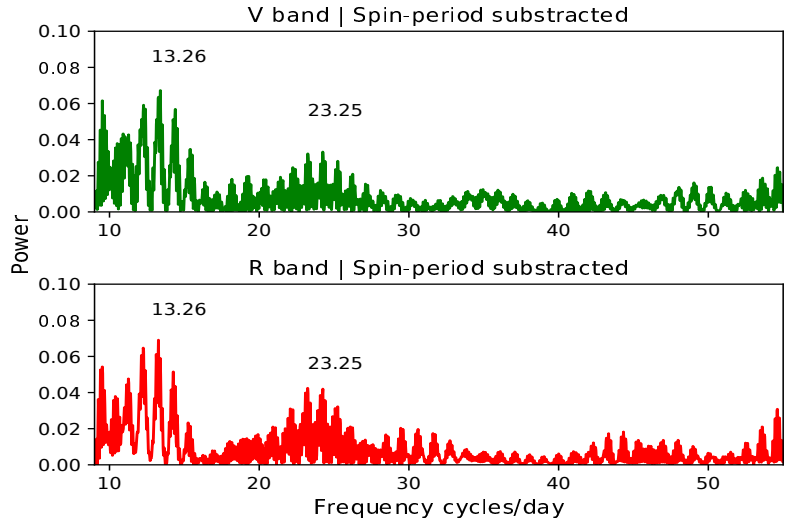}
    \caption{Power spectrum
of the residuals, after removing the spin period signal from the photometric data. }
    \label{fig:PS-foto-substracted}
\end{figure*}

The results from the R  band (middle panel) show the same signals as the V band. In the bottom panel, we show the combined analysis using both the V and R bands. The results are the same as before, where no significant power signals are found for the orbital and beat periods. We note here that the 23.25 signal did not yield a FAP below the 0.01 threshold. To probe for its legitimacy and make an attempt of unmasking new signals, we followed \citet{patterson:2004}, by fitting a sine wave with a periodicity corresponding to the predominant 37.47 cycles/day signal, and then subtracting the fitted sinusoid from the photometric time series. We proceeded to make a new Lomb-Scargle search of the residuals as exhibited in Figure~\ref{fig:PS-foto-substracted}. It can be observed that after subtracting the spin cycle, both of the 13.26 and 23.25 signals remain present, with their recalculated FAPs below the 0.01 cutoff value. This result is different from that obtained by \citep{patterson:2004}, whose power search yields a weak bump at 16 cycles/day after subtracting the spin and beat modulations; although this detection weakens when they study the long-term behaviour of their data. \par
Before folding the photometric time series by the modulations found in the power spectra, we made an analysis to check whether the V and R data are correlated. For this purpose we used a code\footnote{See https://bitbucket.org/pedrofigueira/bayesiancorrelation/src/master/} developed by \citet{figueira:2016}, which implements a Bayesian approach to produce the probability distribution of the correlation coefficient $\rho$. The V and R data yielded a distribution with a mean value of 0.367, a standard deviation of 0.034, and with a 95\% credible interval of [0.298,0.431]. The lower limit of the 95\% credible interval  is well above $\rho=0$, which establishes sufficient confidence in the correlation.\par
\begin{figure*} 
    \includegraphics[width=\columnwidth, height=12cm]{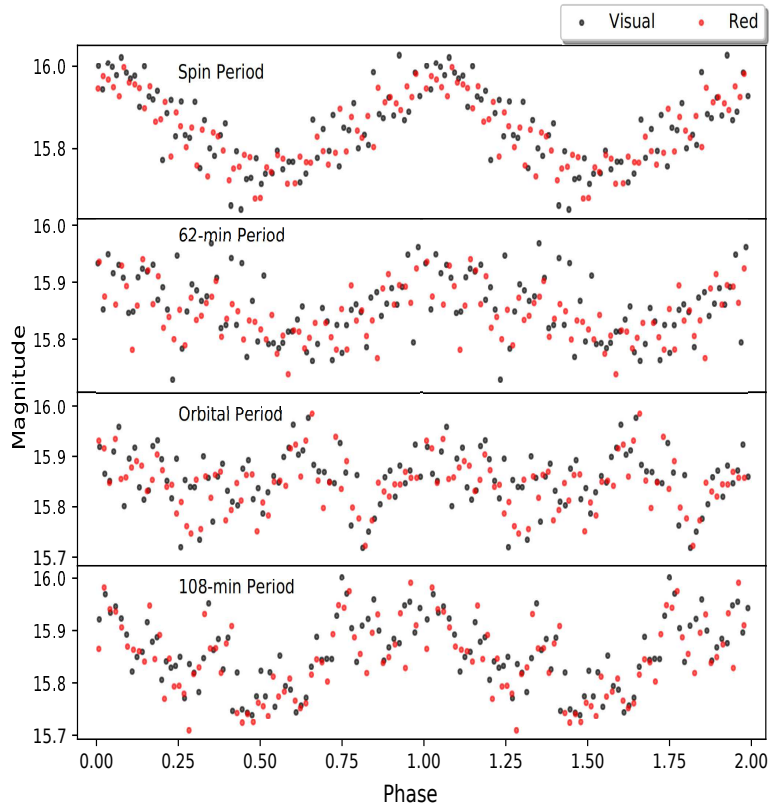}
	\caption{Photometric data folded by the period signals found in the power spectrum search. The data have been averaged into 60 phase bins. The $HJD_{0}$ was selected manually to begin each modulation at its highest, except for the orbital period for which we used the $HJD_{0}$ found in \citet{segura:2020}.}
    \label{fig:curva-doblada-fotometria}
\end{figure*}

In Figure~\ref{fig:curva-doblada-fotometria} we show from top to bottom, the photometric data folded by the associated spin period, the 62-min period, the associated orbital period and  the 108-min period. To reduce the influence of noise we have averaged the data into 60 phase bins.
Folding by the spin period clearly depicts a strong modulation, as expected from the power spectrum analysis.
The 62-min and 108-min folded data show a slightly noisier yet clearly perceptible sinusoidal oscillation. It is worth noting that the 62 min signal corresponds to the beat period between the spin and 108-min periods. Finally, as expected from the lack of signal in the power search, folding by the orbital period yields no sinusoidal modulation for the photometry. 



\subsection{Spectroscopic data}
\label{sec:ps-spec}

\subsubsection{{\sc gau2} Option}
\label{sec:ps-gau2}
\begin{figure*} 
    \includegraphics[width=\columnwidth, height=7cm]{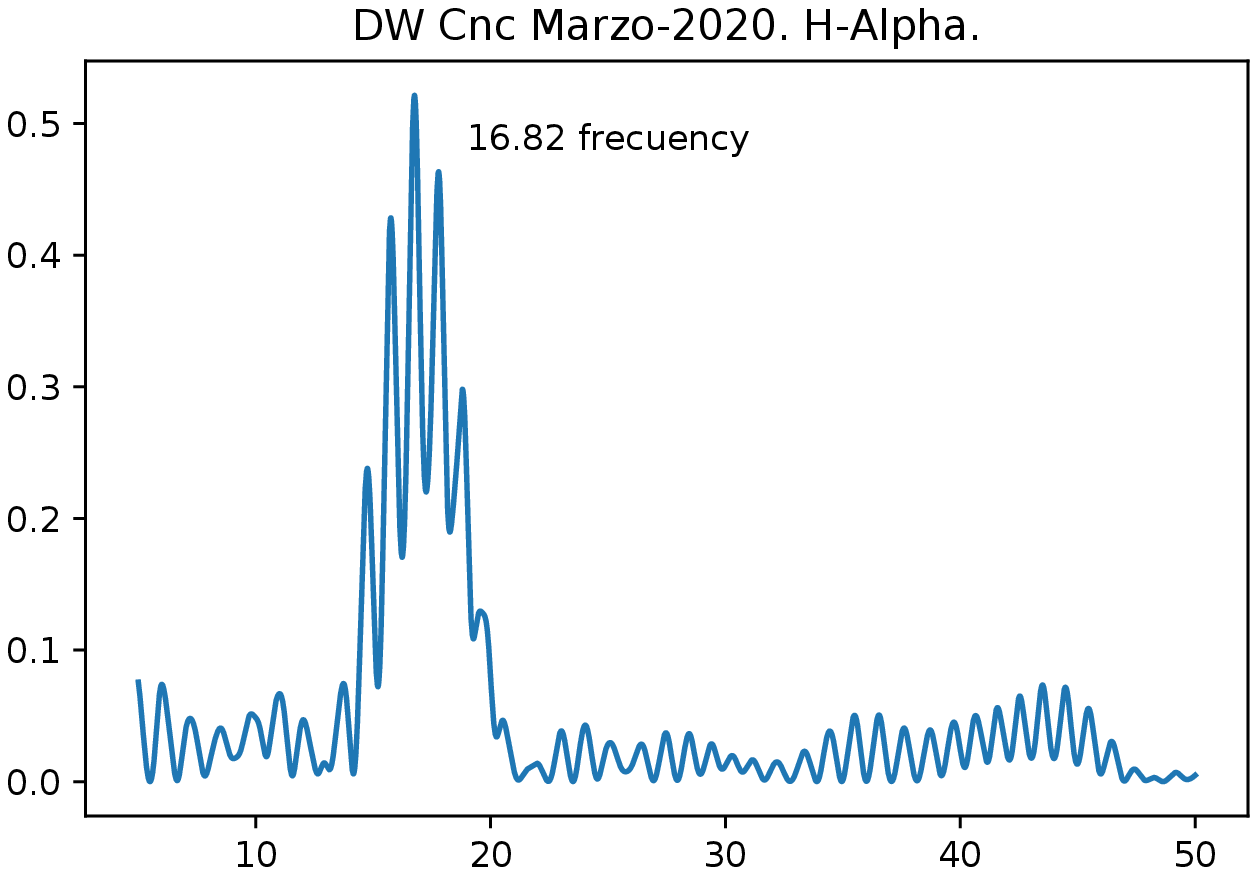}
    \includegraphics[width=\columnwidth, height=7cm]{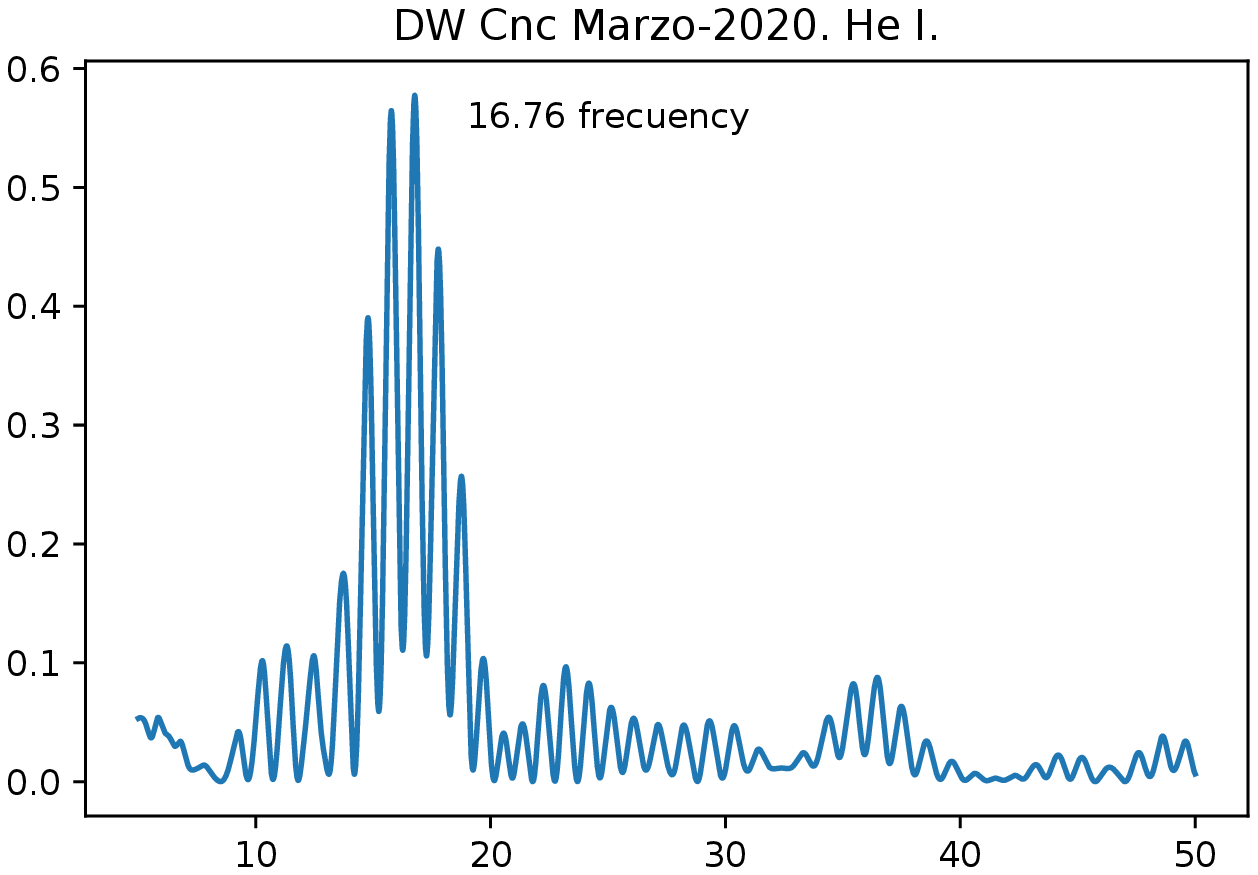}
	\caption{Top: Power spectra of  $H\alpha$ using {\sc gau2}. Bottom: Power spectra of the He I $\lambda 5876$ \AA ~emission line using {\sc gau2}. See text for further discussion.}
    \label{fig:PS-espectroscopia}
\end{figure*}

As described in Section~\ref{sec:radvel}, we performed the measurement of the radial velocities using the {\sc gau2} convolution option. We then proceeded to implement the Lomb-Scargle power search on the convolved data, which is shown in Figure~\ref{fig:PS-espectroscopia}. The upper panel is the analysis of H$\alpha$, which shows a peak frequency at 16.82 cycles/day, corresponding to the 86 min orbital period. No significant power signals for the spin  and beat periods were found. In the lower panel we repeated the analysis for He~I~$\lambda 5876$~{\AA}, which shows a peak frequency at 16.76~cycles/day, associated also with the 86 min orbital period. Again, the spin and beat period signals are not present. The mean value of the orbital period for both lines is 85.76~$\pm ~0.15$ min. This value is, within the errors, compatible with that measured by \citet{segura:2020}. Since we observed a smaller number of spectra than the previous authors, we have a larger error. Therefore we will adopt their value of $P_{orb}=~$86.10169~$\pm$~0.00031 min, throughout this paper. \par
\begin{figure*} 
    \includegraphics[width=\columnwidth, height=7.5cm]{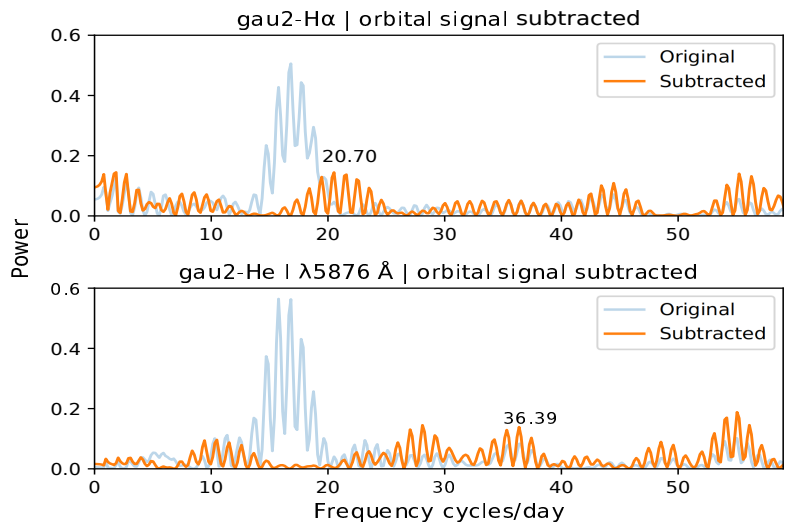}
    \caption{The upper panel shows the power spectrum
of the residuals of the {\sc gau2} H$\alpha$ radial velocity data, after removing the orbital period signal. The lower panel shows the same for He I $\lambda 5876$ \AA. On each panel, the solid orange line represents the power search performed on the residuals. For comparison, we superposed the power spectrum of the original data, plotted as the faint solid blue line. }
    \label{fig:PS-spec-subtracted-gau2}
\end{figure*}

Following the methodology used for the photometric data, described in Section~\ref{sec:photometricdata}, we subtracted the conspicuous orbital signal from both the $H\alpha$ and He I $\lambda~5876$ \AA~data sets and performed a power search on the residuals (shown in Figure~\ref{fig:PS-spec-subtracted-gau2}). The subtracted data of $H\alpha$ (top panel), as with the {\sc dgau} option (see Section~\ref{sec:ps-dgau}), shows the appearance of a weak picket fence around 20.70 cycles/day, with an overly high FAP value of 0.92. On the other hand, the residuals from the He~I~$\lambda 5876$~\AA~ data (bottom panel) show a signal at 36.39 cycles/day, a frequency comparable to the spin modulation, but its 0.97 FAP value puts its legitimacy in doubt.

\begin{figure*} 
    \includegraphics[width=\columnwidth, height=19cm]{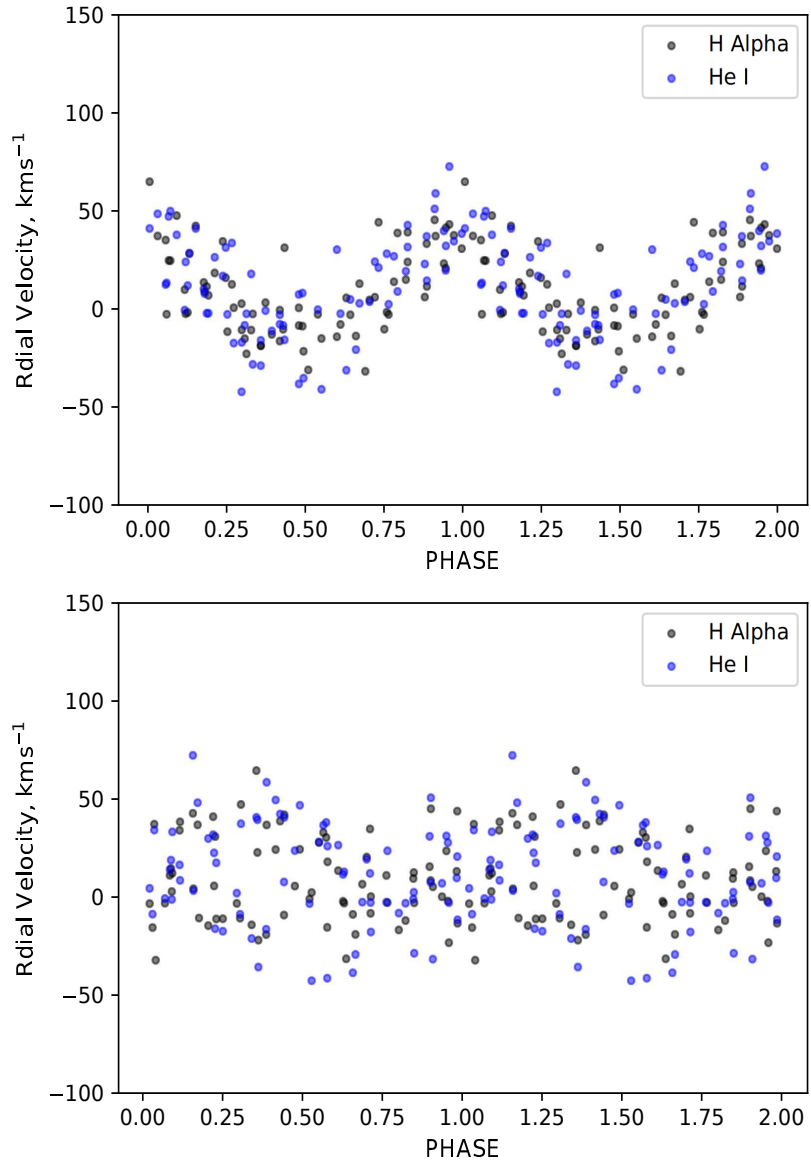}
	\caption{Folded spectroscopic data. The top panel is folded by the orbital period. The bottom panel is folded by the spin period.}
    \label{fig:curva-doblada-espectroscopia}
\end{figure*}

We folded the data by the orbital and spin periods, as shown in Figure \ref{fig:curva-doblada-espectroscopia}. The upper panel exhibits a clear modulation when the data is folded by the orbital period. However, folding by the spin period (bottom panel) shows no evident indication of a periodic signal. \par

\subsubsection{{\sc dgau} Option}
\label{sec:ps-dgau}
\begin{figure*} 
    \includegraphics[width=\columnwidth, height=8cm]{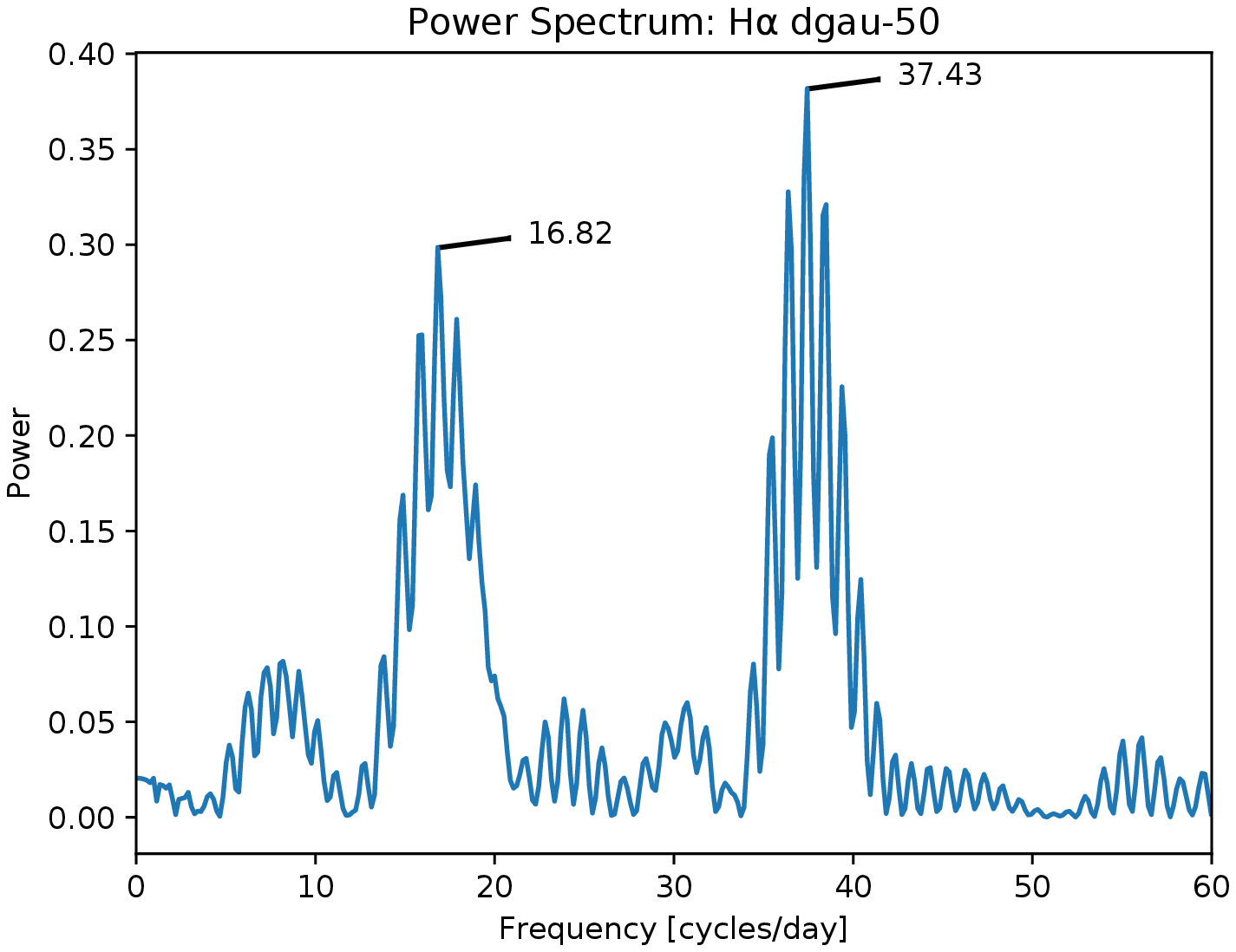}
    \includegraphics[width=\columnwidth, height=8cm]{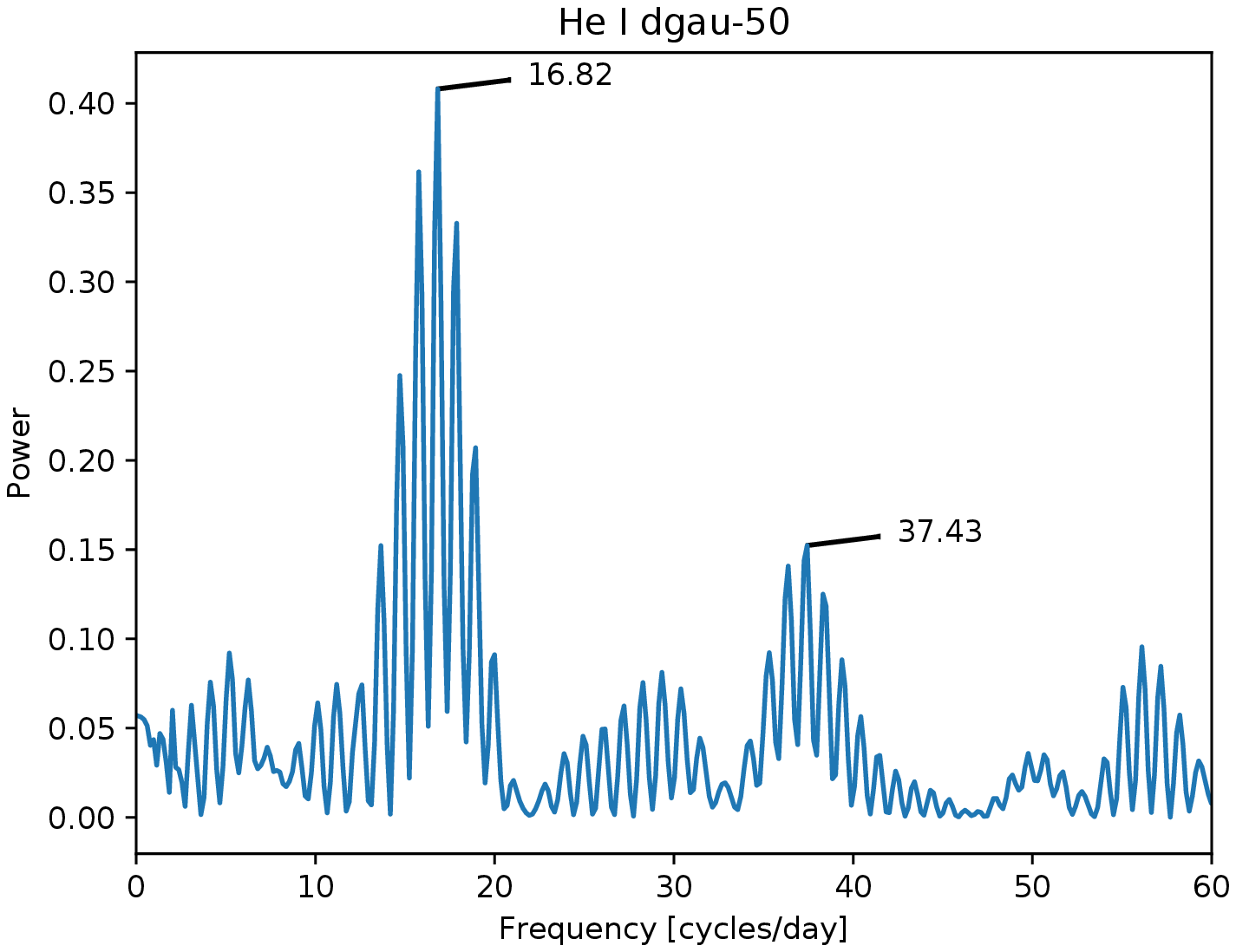}
	\caption{Power spectra of the $H\alpha$ (top) and He I $\lambda 5876$ \AA~(bottom) emission lines, using the dgau option with a Gaussian width of 50 pixels (see text). }
    \label{fig:PS-dgau50}
\end{figure*}

Following \citet{patterson:2004} we measured the radial velocities by convolving the emission lines with the derivative of a Gaussian ({\sc dgau} option) of 83.8 \AA~(50 pixels) of width. As can be observed in Figure \ref{fig:PS-dgau50}, performing a power search on the data yielded the spin and orbital periods for $H\alpha$ (Top panel). The He I $\lambda 5876$ \AA~emission line (bottom panel) shows a clear signal of the orbital period and a weak power peak at the spin period. The computed FAP for this weaker spin peak yielded 0.94, rendering it as questionable signal for He I $\lambda 5876$ \AA. This results are in agreement with those obtained by \citet{patterson:2004} in their spectroscopic power search, where they found the orbital and spin periods.
\par

\begin{figure*} 
    \includegraphics[width=\columnwidth, height=7.5cm]{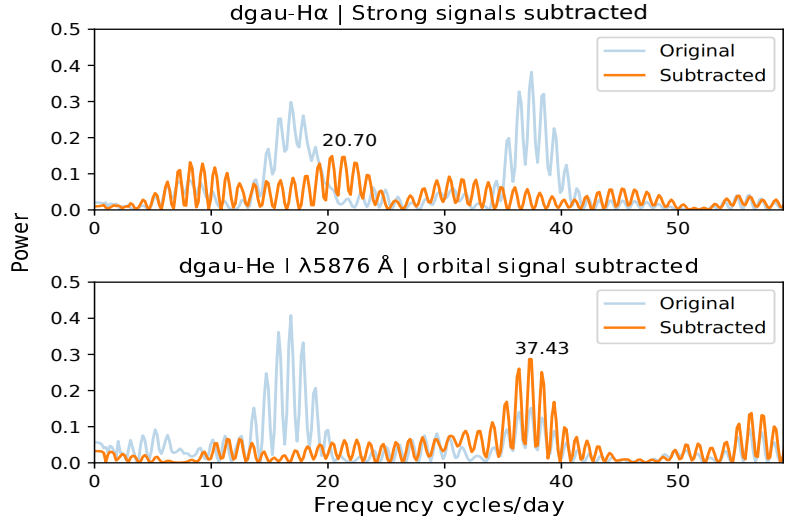}
    \caption{The upper panel shows the power spectrum
of the residuals of the {\sc dgau} H$\alpha$ radial velocity data, after removing the spin period and orbital period signals. The lower panel shows the same for He I $\lambda 5876$ \AA, after subtracting the orbital period. On each panel, the solid orange line represents the power search performed on the residuals. For comparison, we superposed the power spectrum of the original data, plotted as the faint solid blue line. }
    \label{fig:PS-spec-subtracted}
\end{figure*}

 As before, we subtracted the strong modulations from the {\sc dgau} radial velocity data sets of both emission lines, and performed a new Lomb-Scargle search on the residuals. After subtracting the orbital signal from the He I $\lambda 5876$ \AA~data (bottom panel of Figure~\ref{fig:PS-spec-subtracted}), the spin signal peak in the power search of its residuals shows a considerable increase, showing a new FAP value of 0.01 and establishing better reliability upon this modulation. After subtracting both the orbital and spin signals from the H$\alpha$ data (top panel of Figure~\ref{fig:PS-spec-subtracted}), its residuals hint at a weak surge of the 20.70 cycles/day signal, consistent with the {\sc gau2} option in Section~\ref{sec:ps-gau2}. However, the FAP of this signal yields 0.93, a value far too high. Nonetheless, we also detected this signature in the wavelet analysis in Section~\ref{sec:wavelet}, indicating that its presence could be real.
 
\par
\begin{figure*} 
    \includegraphics[width=\columnwidth, height=12cm]{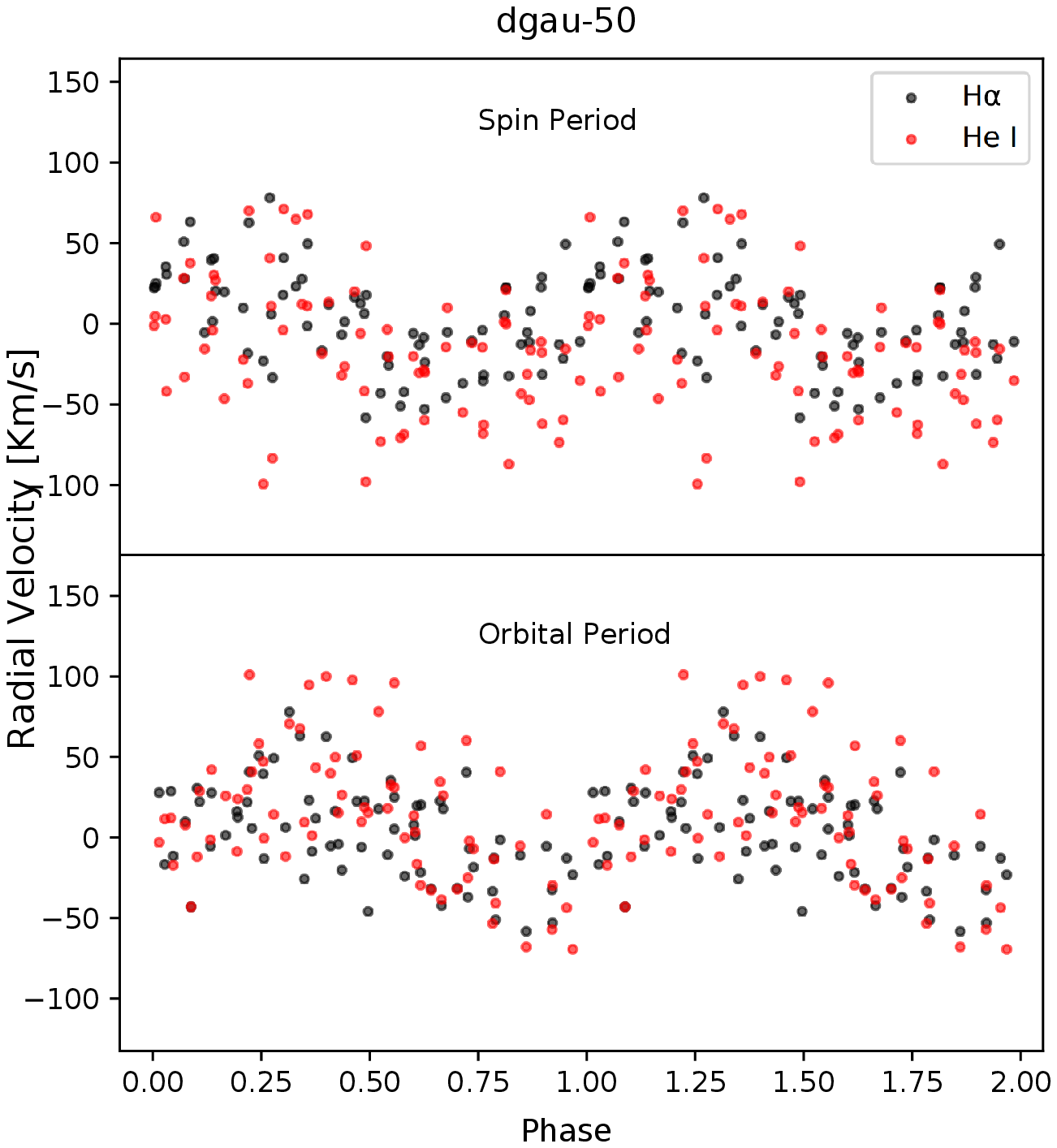}

	\caption{Radial Velocity data of H$\alpha$ and and He I $\lambda 5876$ \AA, folded by the spin (top panel) and orbital (bottom panel) periods found in the Power Spectrum search. }
    \label{fig:rv-dgau50}
\end{figure*}

In Figure \ref{fig:rv-dgau50} we folded the data by the spin and orbital periods, where both signals show clear modulations of the {\sc dgau} radial velocities.

\section{Wavelet Transform Analysis}
\label{sec:wavelet}
The wavelet transform is a method that applies the convolution of the signal with a set of wavelets to map the  variations  occurring in both the time and frequency domains \citep[See][and references therein for a detailed formulation of the method.]{bravo:2014}. The wavelet map, otherwise known as the scalogram, is a useful tool that allows the detection of the scales (or frequencies) that contribute most to the total energy of the signal \citep[e.g.][]{benitez:2010}. \par
Following \citet{bravo:2014} and \citet{deLira:2019}, we performed a wavelet transform analysis of the photometric lightcurve and the spectroscopic radial velocity data. We applied the continuous 6th order Morlet wavelet transform, implemented from the Python ObsPy package\footnote{Available at: https://github.com/obspy/obspy/wiki/} \citep{obspy:2015}. In Figures~\ref{fig:wt-ha-photov}--\ref{fig:wt-he-dgau} we show the local and global power spectra of our data sets. As explained by \citet{bravo:2014}, the local spectrum depicts the energy distribution in time-frequency space, and the global spectrum is obtained by time integration of the local map.

\subsection{Photometry}
\begin{figure*} 
    \includegraphics[width=1\columnwidth, height=9cm]{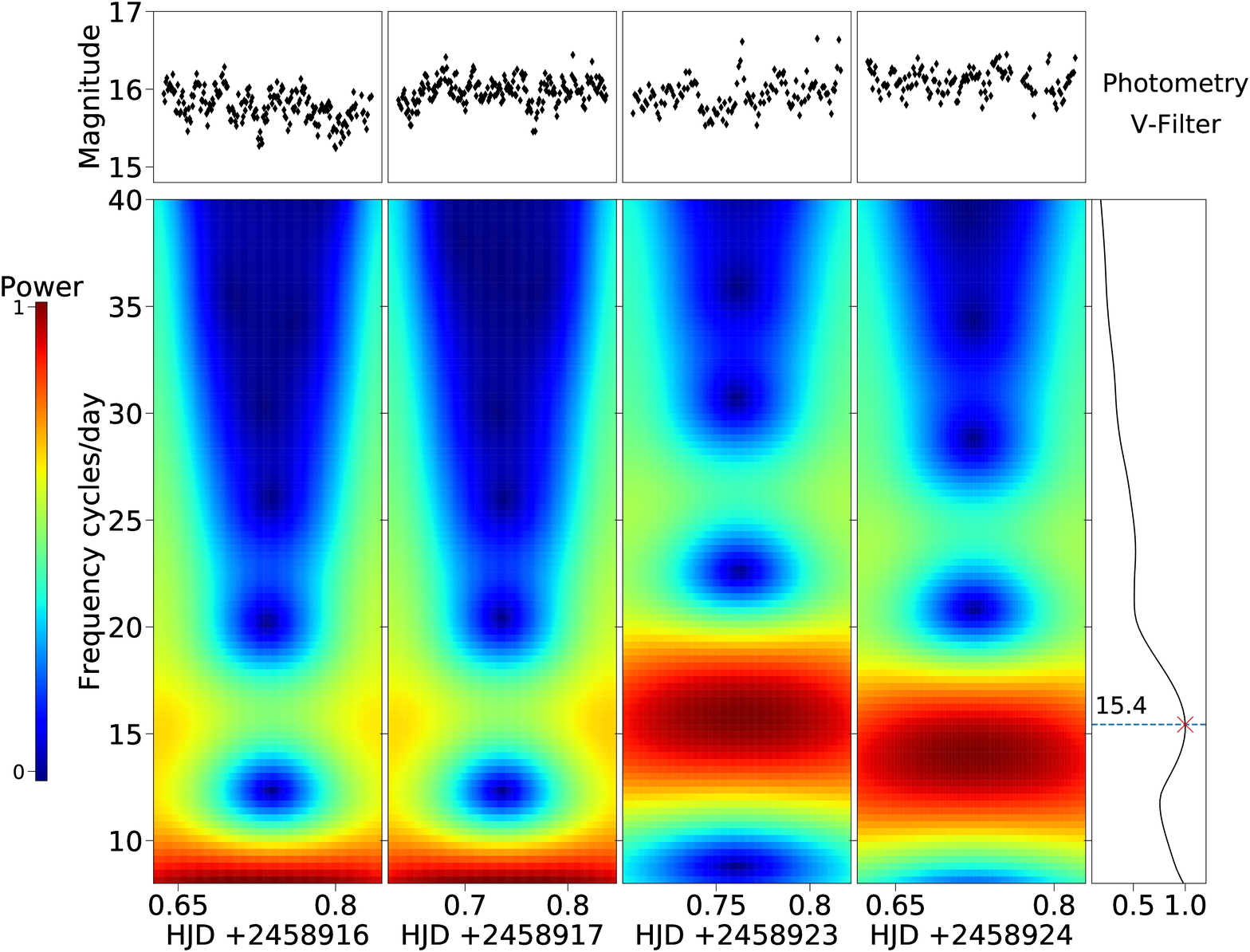}

	\caption{Scalogram of the V-band photometric data. The global spectrum is exhibited in the utmost right panel, while the local spectrum appears on the panels below the data of each night. An orbital cycle signal ($\sim$16 cycles/day) is present throughout all 4 nights of observations. See text for further discussion.}
    \label{fig:wt-ha-photov}
\end{figure*}

The scalogram of the V-band (Figure~\ref{fig:wt-ha-photov}) shows a predominant signal at 15.4 cycles/day. This signal is detected in all four nights, and its relative intensity increases on the third (HJD +2458923) and fourth night (HJD +2458924). This signature is consistent to the orbital cycle, which was absent in the Lomb-Scargle power spectrum. 
\par
\begin{figure*} 
    \includegraphics[width=1\columnwidth, height=9cm]{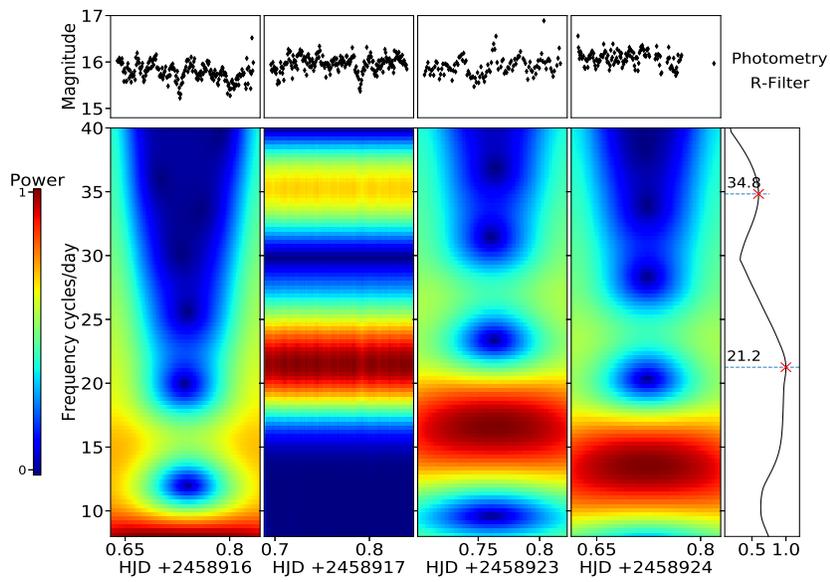}

	\caption{Scalogram of the R-band photometric data. The global spectrum is exhibited in the utmost right panel, while the local spectrum appears on the panels below the data of each night. The second night shows the presence of the spin period and a $\sim$21 cycles/day signal (consistent with the spin-orbit beat period). The rest of the nights are dominated by the orbital cycle. See text for further discussion.}
    \label{fig:wt-ha-photoR}
\end{figure*}

The R-Band analysis (Figure~\ref{fig:wt-ha-photoR}) shows a similar structure to that exhibited by the V-band scalogram, except for the second night (HJD +2458917), which yields a strong signal at $\sim$21 cycles/day; this frequency is comparable to the 20.60 cycles/day modulation found by \citet{patterson:2004}, corresponding to the beat period between the spin and orbital cycles. We note that we did not detect this signal in the photometric power Lomb-Scargle analysis (see Section~\ref{sec:photometricdata}), but it did appear after subtracting the stronger signals in the spectroscopic data in Section~\ref{sec:ps-spec}. A secondary signal peaking at $\sim$35 cycles/day (possibly associated with the spin period) is also visible during the second night (HJD +2458917).

\subsection{Spectroscopy}
We performed the wavelet analysis on the radial velocity data sets obtained both from the {\sc gau2} and {\sc dgau} convolution methods. We now proceed to describe this results. \par

\subsubsection{{\sc gau2 }Option}
\begin{figure*} 
    \includegraphics[width=1\columnwidth, height=9cm]{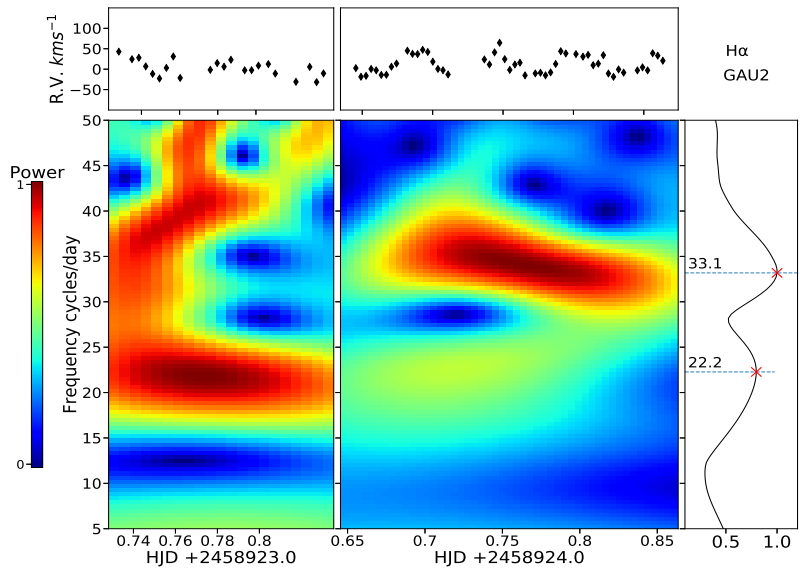}

	\caption{Scalogram of the $H\alpha$~{\sc gau2} spectroscopic data. The global spectrum is exhibited in the utmost right panel, while the local spectrum appears on the panels below the data of each night. The first night shows  prominent signals at $\sim$37 cycles/day and $\sim$22 cycles/day. The latter signal also appears during the second night but is surpassed in intensity by a $\sim$33 cycles/day signature. See text for further discussion.}
    \label{fig:wt-ha-gau2}
\end{figure*}

The $H\alpha$ {\sc gau2}~scalogram in Figure~\ref{fig:wt-ha-gau2}, shows a broad power peak extending from $\sim$20 to $\sim$39 cycles/day at the outset of the first night (HJD +2458923.0). Such broad power peak narrows down onto two localized peaks, of which the most prominent and persistent shows a midpoint at a frequency of $\sim$22 cycles/day; a signal consistent with that also found in the photometric Lomb-Scargle analysis of $\sim$23 cycles/day. The second night of this scalogram (HJD +245894) displays a strong signal at $\sim$33 cycles/day, which we consider to be a possible alias of the spin modulation. The $\sim$22 cycles/day signature is also present during the second night but with a decrease in relative intensity.\par

\begin{figure*} 
    \includegraphics[width=1\columnwidth, height=9cm]{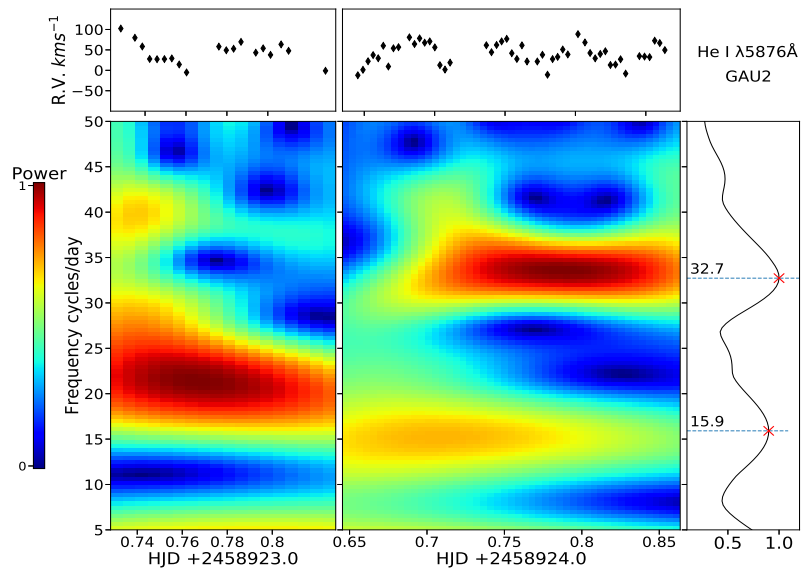}

	\caption{Scalogram of the He I $\lambda 5876$\AA~{\sc gau2} spectroscopic data. The global spectrum is exhibited in the utmost right panel, while the local spectrum appears on the panels below the data of each night. The first night shows a dominant signal at $\sim$22 cycles/day and a second one at $\sim$40 cycles/day. The second night is dominated by a $\sim$33 cycles/day signature, accompanied by a persistent secondary signal at $\sim$16 cycles/day. See text for further discussion.}
    \label{fig:wt-he-gau2}
\end{figure*}

The He I $\lambda5876$~\AA~{\sc gau2} scalogram, exhibited in Figure~\ref{fig:wt-he-gau2}, shows a prominent signal that persists throughout the first night (HJD +2458923) at $\sim$22~cycles/day, in agreement with the $H\alpha$ {\sc gau2} data. During this first night a secondary signal appears at $\sim$40 cycles/day, which gradually shifts to $\sim$37 cycles/day (related to the spin period) as the night progresses. During the second night (HJD +2458924) we see an overall shift towards smaller frequency values, displaying a strong signal at $\sim$33 cycles/day and a slightly milder yet very persistent signal at $\sim$16 cycles/day (consistent with the orbital period). \par

\subsubsection{{\sc dgau }Option}
\begin{figure*} 
    \includegraphics[width=1\columnwidth, height=9cm]{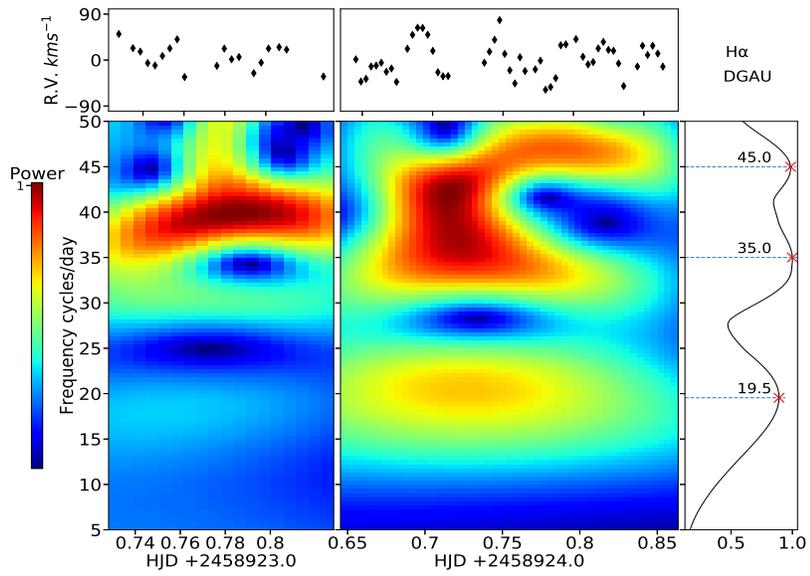}

	\caption{Scalogram of the $H\alpha$ {\sc dgau} spectroscopic data. The global spectrum is exhibited in the utmost right panel, while the local spectrum appears on the panels below the data of each night. The first night shows a strong signal, consistent with the spin cycle, at $\sim$38 cycles/day. During the second night the spin cycle signal is also present although with a broader structure; a secondary signature also appears this night at $\sim$20 cycles/day, which is consistent with the spin-orbit beat period. See text for further discussion.}
    \label{fig:wt-ha-dgau}
\end{figure*}

Figure~\ref{fig:wt-ha-dgau} depicts the $H\alpha$ 
{\sc dgau} scalogram, showing during the first night
(HJD +2458923), a conspicuous signal at the expected spin period of $\sim$38 cycles/day. In the second night (HJD +2458924), the signal observed the previous night is still visible, but its structure considerably broadens in frequency, eventually adopting a two-pronged shape, whose contribution reflects on the loss of the 38 cycles/day signal in the global spectrum. The second night also shows a clear signal at $\sim$20 cycles/day, related to the 70 min beat period. \par

\begin{figure*} 
    \includegraphics[width=1\columnwidth, height=9cm]{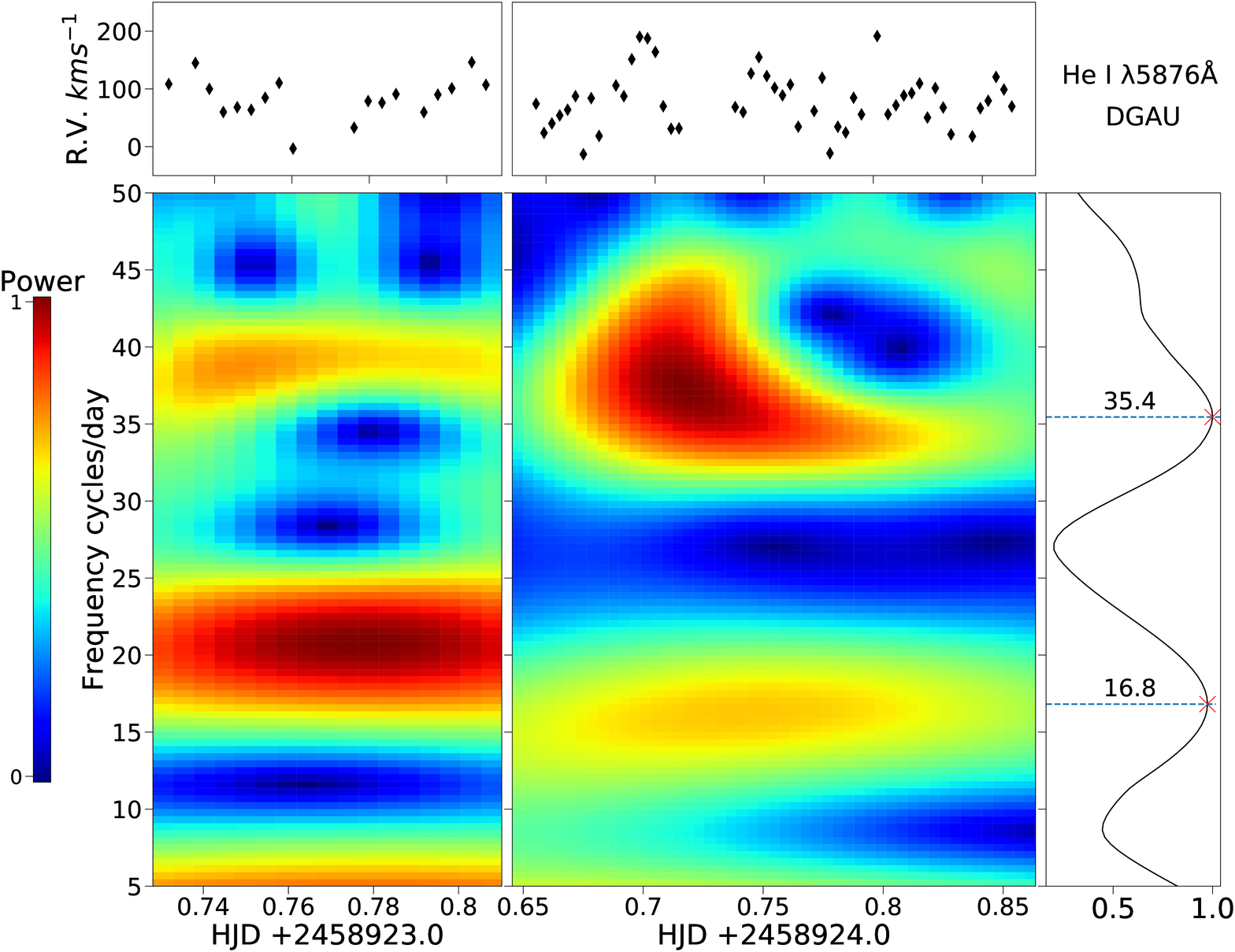}

	\caption{Scalogram of the He I $\lambda 5876$\AA~{\sc dgau} spectroscopic data. The global spectrum is exhibited in the utmost right panel, while the local spectrum appears on the panels below the data of each night. The first night shows a strong and persistent detection at the $\sim$20 cycles/day beat signal. Another signature is evident at the $\sim$38 cycles/day spin signal. The latter is also present during the second night, although its structure considerably broadens. The second night shows also the appearance of a persistent signal of the orbital cycle at $\sim$16 cycles/day. See text for further discussion.}
    \label{fig:wt-he-dgau}
\end{figure*}

The He I $\lambda5876$~\AA~{\sc dgau} scalogram, in Figure~\ref{fig:wt-he-dgau}, displays a strong signal at $\sim$20 cycles/day that persists all throughout the first night (HJD +2458923), and a weaker signature at $\sim$38 cycles/day, consistent with the spin modulation. On the second night ((HJD +2458924) the spin cycle signal becomes enhanced and its structure considerably broadens. This night also shows the appearance of a secondary detection at $\sim$16 cycles/day, related to the spin modulation, which persists the whole night.

\section{Doppler Tomography}
\label{sec:tom}
\begin{figure*}
\centering
	\includegraphics[angle=0,height=8cm,width=1.0\columnwidth]{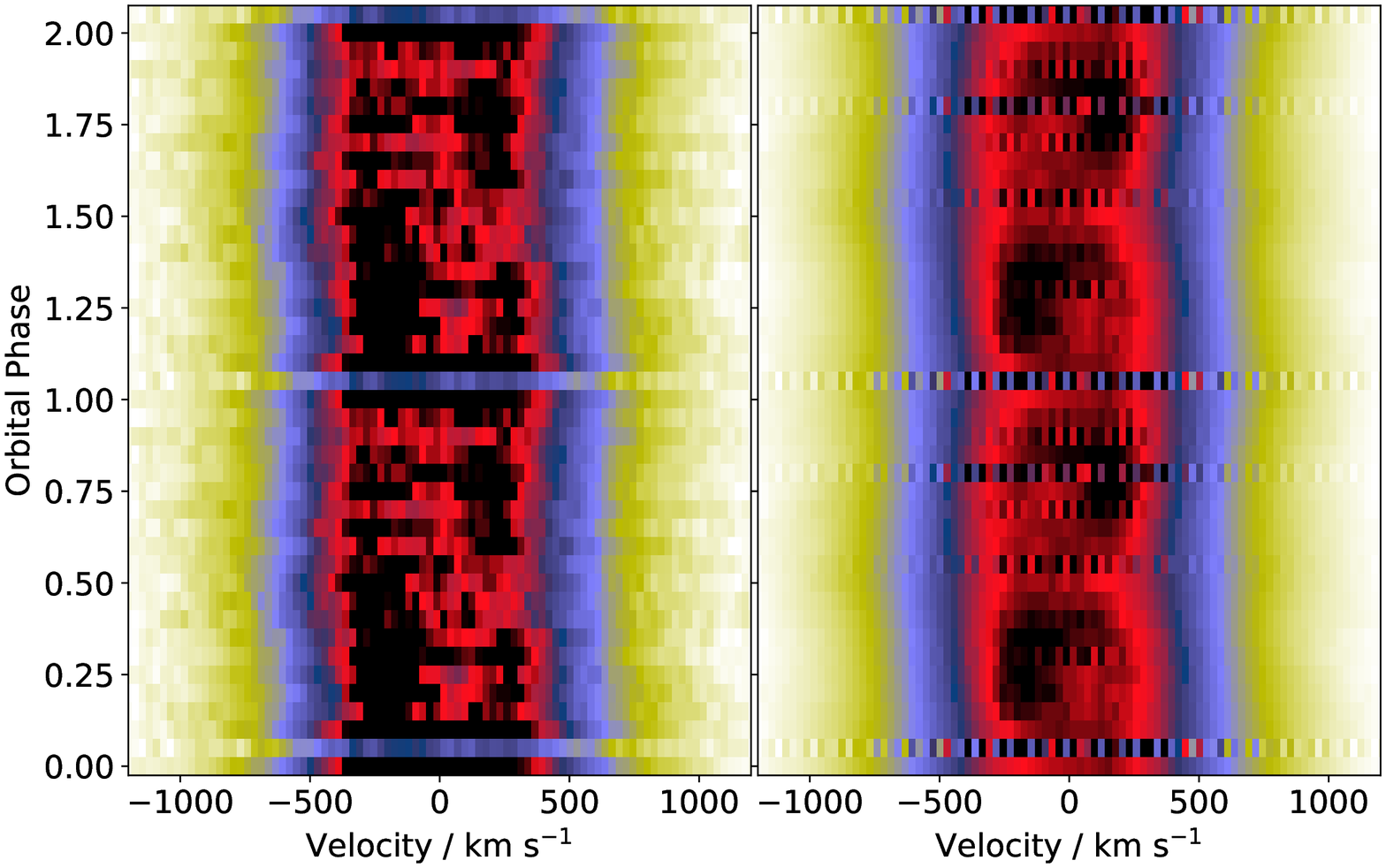}
	\includegraphics[angle=0,height=8cm,width=1.0\columnwidth]{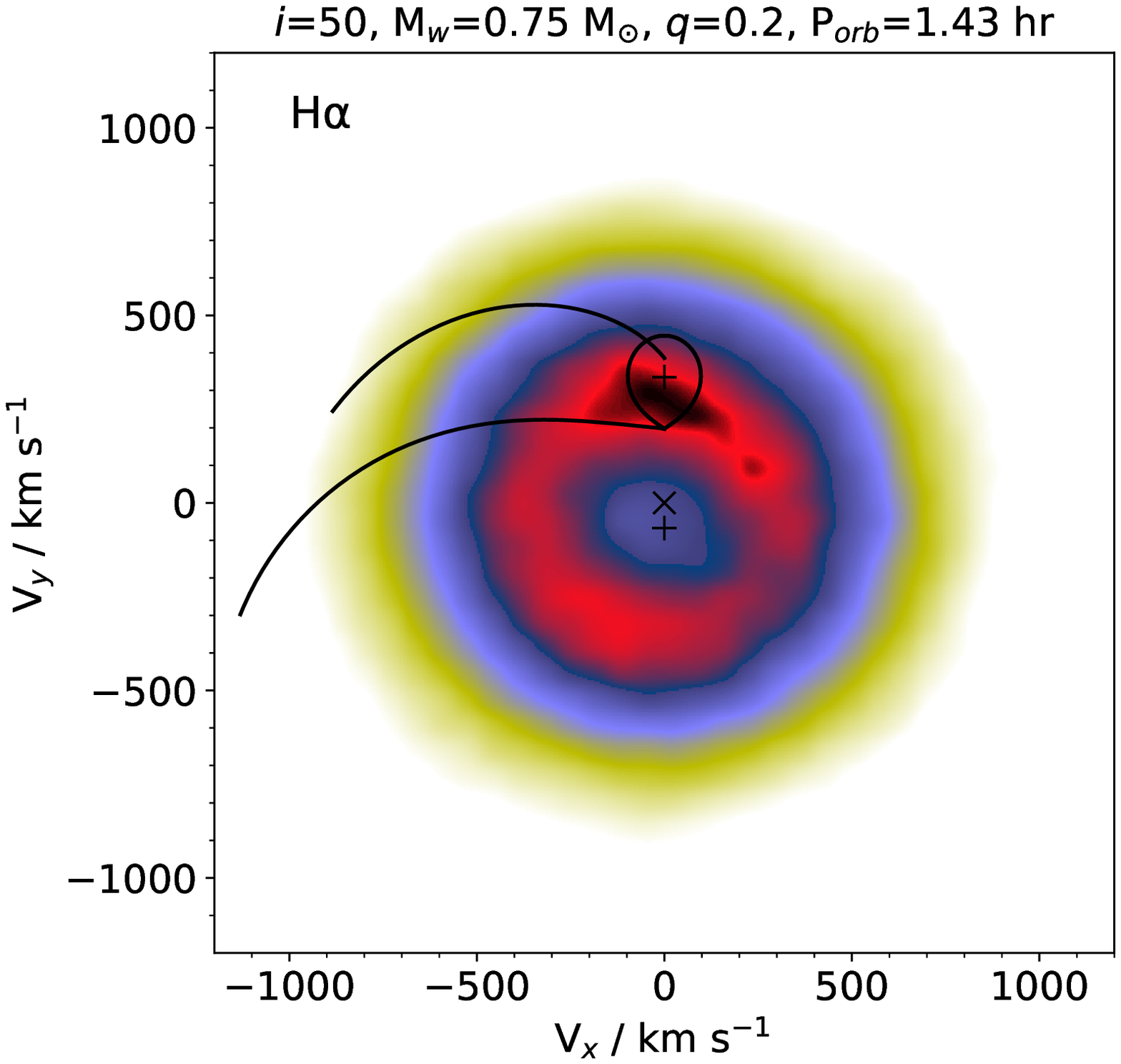}
\caption{Trail spectra and Doppler Tomography of the $H\alpha$ emission line. The relative emission intensity is shown in a scale of colours, where the strongest intensity is represented by black, followed by red, then blue, and finally yellow. The cross markings represent (from top to bottom) the position of the secondary, the centre of mass and the primary component. The Roche lobe of the secondary is depicted around its cross. The Keplerian and ballistic trajectories of the gas stream are marked as the upper and lower
curves, respectively.}	
	
\label{dopmap-ha} 
\end{figure*}

\begin{figure*}
\centering
	\includegraphics[angle=0,height=8cm,width=1.0\columnwidth]{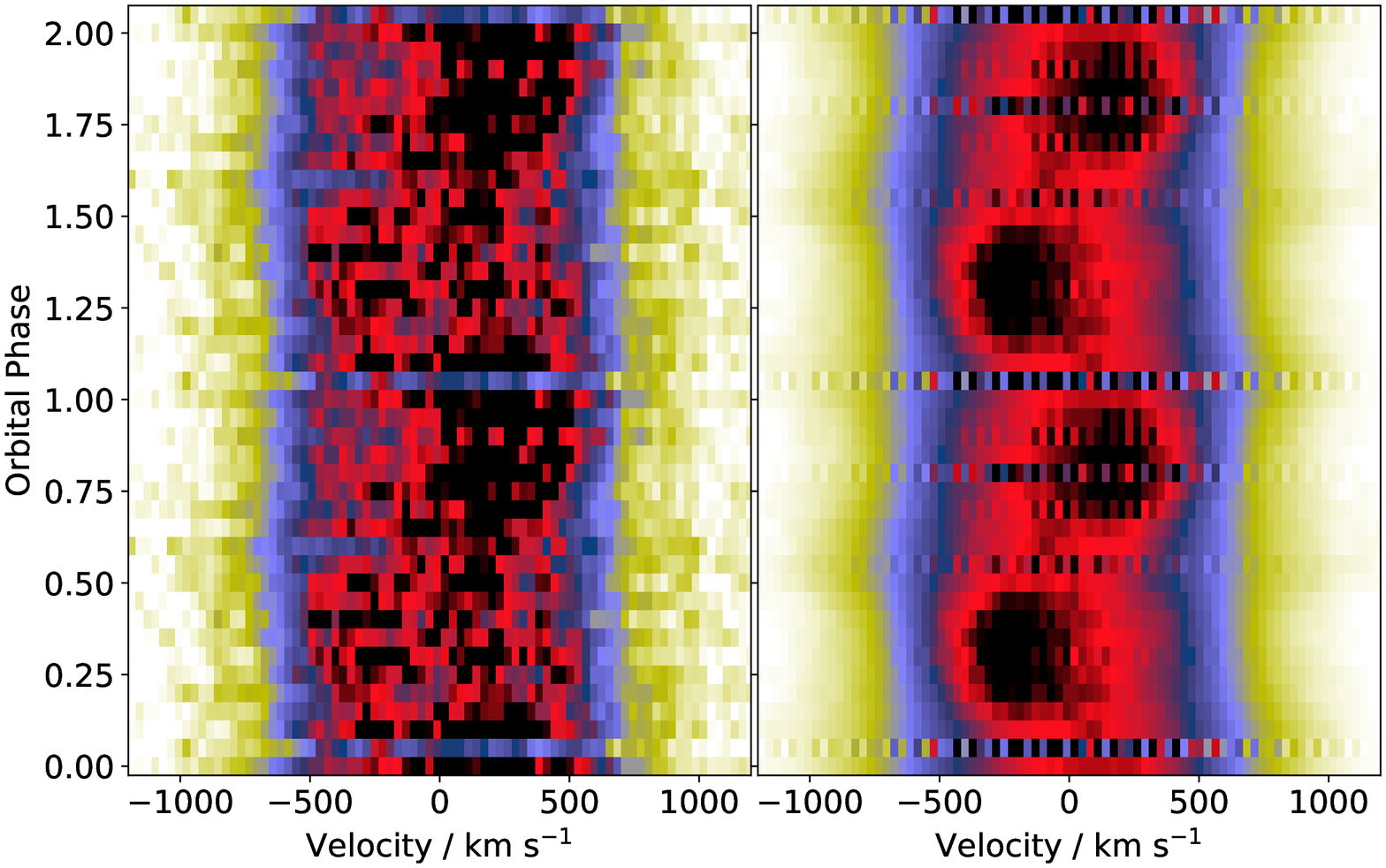}
	\includegraphics[angle=0,height=8cm,width=1.0\columnwidth]{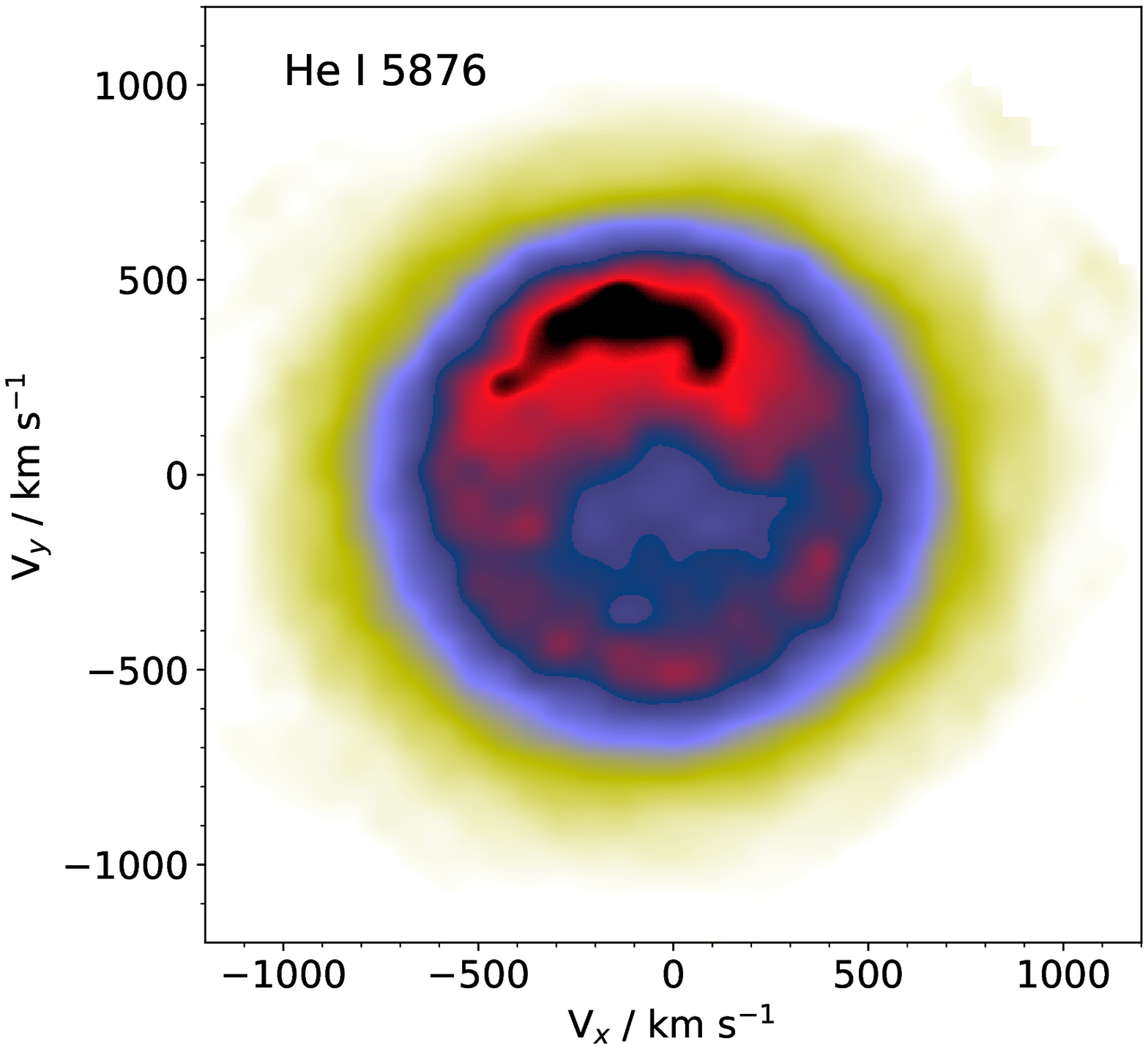}
\caption{Trail spectra and Doppler Tomography of the He I~$\lambda$ 5876~\AA\ emission line. The relative emission intensity is shown in a scale of colours, where the strongest intensity is represented by black, followed by red, then blue, and finally yellow. The cross markings represent (from top to bottom) the position of the secondary, the centre of mass and the primary component. The Roche lobe of the secondary is depicted around its cross. The Keplerian and ballistic trajectories of the gas stream are marked as the upper and lower
curves, respectively. }	
	
\label{dopmap-he} 
\end{figure*}

Doppler Tomography is an indirect imaging technique developed by \citet{marsh:88}. It produces two-dimensional mappings of the emission intensity in velocity space of the accretion disc, using the phase-resolved profiles of the spectral emission lines. We obtained the Doppler Tomography of the $H\alpha$ and of the He I  $\lambda 5876$~\AA~emission lines, using a Python wrapper \footnote{Available at \url{https://github.com/Alymantara/pydoppler}} \citep{pydoppler:2021} of the original {\sc fortran} routines published by \citet{spruit:1998} within an {\sc idl} environment. In the top left panel of Figure \ref{dopmap-ha} we show the observed trailed spectra of $H\alpha$, while the reconstructed trailed spectra appears in the top right panel; the tomogram is displayed in the bottom panel. With the same layout, the trailed spectra and tomography of  He I~$\lambda$ 5876~\AA~ are exhibited in Figure~\ref{dopmap-he}. The parameters used to plot the features in the tomograms are as follows: an inclination of $i=50^{\circ}$; a  value of the mass of the primary star of $M_w$=0.75 $M_{\odot}$, consistent with the average mass for white dwarfs in CVs below the period gap \citep{knigge:2006}; a mass ratio $q=0.2$, estimated following \citet{echevarria:1983}; and an orbital period of $P_{orb}=86.10$ minutes \citep{patterson:2004,rodriguez:2004,segura:2020}. We now proceed to describe the results obtained for each emission line. 
\subsection{$H\alpha$}
The $H\alpha$ observed trailed spectra displays a complex behaviour. From orbital phase 0.0 to $\sim$0.10 it shows a single peaked structure. In the interval from 0.10 to 0.30, the trailed spectra displays a double-peaked profile. After this interval, the profile briefly becomes single peaked, and from 0.35 to 0.55 the blue shifted peak becomes more intense than the red shifted peak. From 0.55 onward the line profile again displays a symmetric double-peaked structure with a brief single-peaked intrusion at $\sim0.8$.\par
The Tomography shows a disc signal in red colour \citep{marsh:88} with a superimposed intense region (in black) at the position of the Roche Lobe of the secondary, which could be caused by emission from a hot spot component \citep[e.g.][]{echevarria:2007}. The disc structure was not detected in the tomography by \citet{segura:2020}, but this finding is in good agreement with the dominant double-peaked structure, characteristic of discs in systems of high inclination \citep{horne:1986}, exhibited by the $H\alpha$ line profiles reported by \citet{rodriguez:2004}.

\subsection{He I 5876 {\AA}}
The He I~$\lambda$ 5876~\AA~trailed spectra show the oscillation of a broad single-peaked profile, consistent with the line profiles of this emission line put forward by \citet{rodriguez:2004}. The Tomography shows a blob-like region of high intensity in the upper quadrants that overlays the position of the Roche Lobe of the secondary, and further extends towards negative velocities, covering the position where the emission coming from the hotspot is expected in velocity space. Note that the position of the blob in the He I~$\lambda$ 5876~\AA~ tomography is consistent with that of the region of maximum intensity observed for $H\alpha$.

\section{Discussion}
\label{sec:discus}

We performed a study of new photometric and spectroscopic observations of DW Cnc, after  the system  recovered from a low state, presumably caused by an episode of low mass transfer from the secondary that inhibited the lighthouse effect \citep{segura:2020}. Photometry from the AAVSO shows that at time of our observations, DW Cnc had already reached a state comparable to that from 1999-2003 reported by \citet{patterson:2004} and \citet{rodriguez:2004}. With this in mind, we performed various analyses to compare the behaviour displayed by DW Cnc before and after experiencing the low state.

Our photometric power search analysis (see Section~\ref{sec:photometricdata}) shows a clear spin-cycle modulation, in agreement with \citet{patterson:2004} and \citet{rodriguez:2004}. Our photometry also exhibits a moderate signal at 108 min. \citet{patterson:2004} found evidence of a similar weak signal at 110.85(9) min; they left this finding as an unsolved problem, which as they put it, \textit{does not seem related to any other clocks in the binary.} Furthermore, we have also detected a new 62 min period signal which, although weak, corresponds to the beat period of the spin and the 108 min signatures. Finding this beat signal suggests that the 108 min period is not caused by spurious effects. Still, the origin of these modulations remains unknown. However, we did not find the 70~min beat period signal, nor the 86~min orbital period after the subtraction of the main spin modulation; these lack of signals is the main contrast in our photometric results from those by \citet{patterson:2004}. We find the same contrast with the modulations found in the X-ray observations of \citet{nea19}, who also detected the spin-orbit beat period and a spin modulation signal.

We also conducted two different power spectrum analyses of the radial velocities of the H$\alpha$ and He~I~$\lambda~5876$~{\AA} emission lines (see Section~\ref{sec:ps-spec}): the first one by convolving the lines with two antisymmetric Gaussians ({\sc gau2}), and the second one by employing the derivative of a Gaussian  ({\sc dgau}) as the convolution function. When employing the same method as \citet{patterson:2004}, i.e. {\sc dgau}, we obtain consistent signals with these authors, detecting both the orbital and spin modulations. On the other hand, the power search on the {\sc gau2} option (which traces the inner regions of the disc), yielded only a strong signal for the orbital period.  This, in a way, is also similar to the results reported by \citet{rodriguez:2004}, who  find both signals when using a \textit{broad} Gaussian but obtain exclusively the  orbital signature when the correlation of the emission line is made with a \textit{narrow} Gaussian.  \par
In Section~\ref{sec:wavelet}, we conducted a wavelet transform analysis which confirmed the presence of some of the signals found in the Lomb-Scargle periodograms. However, and perhaps more remarkably, this analysis also detected a localized signature of the 70~min beat period in the {\sc dgau} scalograms. We note that a weak hint of the beat modulation also appeared in the $H\alpha$ periodograms after subtracting the dominant signals from the spectroscopic data, but we regard this detection with caution given the high FAP yielded by the signature (see Section~\ref{sec:ps-spec}).\par

We implemented a Doppler Tomography study of the binary in Section~\ref{sec:tom}. The trailed spectrum of $H\alpha$ exhibits a double-peaked structure that becomes single-peaked in short intervals and with a changing relative intensity of the peaks. On the other hand, the He I $\lambda~5876$~\AA~trailed spectrum shows a broad single-peaked profile. Both of these trailed spectra are consistent  with the structure of the profiles of the same emission lines reported by \citet{rodriguez:2004}. Moreover, our $H\alpha$ Doppler tomogram displays the presence of the accretion disc, a structure not detected in the tomograms by \citet{segura:2020}, hinting at a possible replenishing of the disc. Our tomography also shows what appears to be a hot spot component for both emission lines, in good agreement with the S-wave, presumably originated at the location of the bright spot, exhibited in the trailed spectra diagrams from \citet{rodriguez:2004}.\par

The similarities with previous studies, listed above, lead us to believe that the system has undergone at least a partial recovery from the low state. In particular, finding a clear indication of the spin period both in our photometry and in the {\sc dgau} option, a signal which avoided detection in \citet{segura:2020}, suggests that the outer disc has undergone enough replenishing to provoke a detectable lighthouse beacon. This is further supported by the tomography showing a clear indication of the accretion disc and a hot spot emission.\par

Nonetheless, the main differences regarding the signals found (and not found) in the Lomb-Scargle power search of the photometry, prevent us from declaring complete recovery of the previous state of DW Cnc. It remains to be seen if the system will fully recover to the condition reported in \citet{patterson:2004}. The results in the present article show some progress, but further observations are still required to see whether if the mechanisms causing the behaviour reported in 2004 require more time to get kick-started and allow detection.

\section{Conclusions}
\label{sec:conclusions}
Photometric and spectroscopic observations of DW Cnc show, to some extent, an agreeable behaviour from that exhibited before experiencing a low state, pointing at a partial recovery of the system. Namely, our photometry yields a strong modulation of the spin period which indicates the reactivation of the lighthouse effect, that eluded detection by \citet{segura:2020}; our analysis also showed a weak unresolved signal at 108 min. Furthermore, when implementing the same methodology as \citet{patterson:2004} to measure the radial velocities of the emission lines, we obtained a consistent result with these authors in the periodogram, where we detect both the orbital and spin periods. 
Furthermore, in agreement with \citet{rodriguez:2004}, we find evidence of the disc structure and hot spot emission in our Doppler Tomography study. Finally, our wavelet transform analysis displays a localized detection of the 70 min beat period.
\par
However, the signatures in the photometric periodograms do not completely match those reported before the low state, by \citet{nea19} and \citet{patterson:2004}, who  find not only the spin cycle, but also  a signal corresponding to the spin-orbit beat period, and even a weak detection of the orbital signal after further treatment of their data. Instead we found a new 62 min period which corresponds to the beat between the  spin and 108~min periods.\par
 We could not find public data that would enable us to replicate our analyses. Therefore, we propose additional observations of the system to assess if it is possible that the mechanisms that gave rise to the signatures exhibited before the low state, require more time to completely rekindle.

\section*{Acknowledgements}

The authors are indebted to DGAPA (Universidad Nacional Aut\'onoma de M\'exico) support, PAPIIT projects IN114917 and IN103120. We thank the staff at the OAN-SPM for facilitating and helping us to obtain our observations. This research made extensive use of {\sc astropy}, a community-developed core Python package for Astronomy \citep{Astropy-Collaboration:2013aa}; Python's SciPy signal processing library \citep[][]{virtanen:2020} and {\sc matplotlib} \citep{Hunter:2007aa}. We also thank the anonymous referee, whose comments substantially helped  improve the content of this paper.

\end{document}